\documentclass[twocolumn,showpacs,eqsecnum,aps]{revtex4}
\usepackage{graphicx}
\usepackage{dcolumn}
\usepackage{amsmath}
\makeatletter
\def\btt#1{\texttt{\@backslashchar#1}}%
\DeclareRobustCommand\bblash{\btt{\@backslashchar}}%
\makeatother
\def\epsi{\varepsilon}
\def\pn{p_N}
\def\pnd{p_N '}
\def\mn{m_N}
\def\md{m_\Delta}
\def\mv{m_V}
\def\pd{p_\Delta}
\def\kd{{\overline k}}
\def\khat{\widehat K}
\def\kperp{k_\perp}
\def\kp2{k_\perp^2}
\def\yd{{y '}}
\newcommand{\yp}[1]{y^{\prime {#1}}}
\begin{document}
\title{\ \ \ \\
       \vspace{-1.7cm}
       \hfill SAGA-HE-176-01, TMU-NT-01-04 \\
                           \ \\
                           \ \\
        Polarized light-antiquark distributions in a meson-cloud model}
\author{S. Kumano}
\homepage{http://hs.phys.saga-u.ac.jp}
\email{kumanos@cc.saga-u.ac.jp}
\affiliation{Department of Physics, Saga University \\
         Saga, 840-8502, Japan}
\author{M. Miyama}
\email{miyama@comp.metro-u.ac.jp}
\affiliation{Department of Physics, Tokyo Metropolitan University \\
         Tokyo, 192-0397, Japan}         
\date{October 6, 2001}
\begin{abstract}
\vspace{0.2cm}
Flavor asymmetry is investigated in polarized light-antiquark distributions
by a meson-cloud model. In particular, $\rho$ meson contributions
to $\Delta \bar u - \Delta \bar d$ are calculated. 
We point out that the $g_2$ part of $\rho$ contributes to
the structure function $g_1$ of the proton
in addition to the ordinary longitudinally polarized
distributions in $\rho$. This kind of contribution becomes important
at medium $x$ ($>0.2$) with small $Q^2$ ($\sim$1 GeV$^2$).
Including $N \rightarrow \rho N$ and $N \rightarrow \rho \Delta$ splitting
processes, we obtain the polarized $\rho$ effects on the
light-antiquark flavor asymmetry in the proton.
The results show $\Delta \bar d$ excess over $\Delta \bar u$,
which is very different from some theoretical predictions.
Our model could be tested by experiments in the near future.
\end{abstract}
\pacs{13.60.Hb, 13.88.+e, 12.39.-x}
\maketitle
\tableofcontents

\section{Introduction}\label{intro}
\setcounter{equation}{0}

Light antiquark distributions are expected to be almost flavor symmetric
according to perturbative quantum chromodynamics (QCD).
The next-to-leading-order effects contribute to the difference between
$\bar u$ and $\bar d$; however, the contribution is tiny as long as
they are estimated in the perturbative QCD region. 
Therefore, it was rather surprising to find the antiquark
flavor asymmetry $\bar u/\bar d$ in Gottfried-sum-rule violation
by the New Muon Collaboration (NMC) \cite{nmc}
and in succeeding Drell-Yan and semi-inclusive measurements
of the CERN-NA51 \cite{na51}, Fermilab-E866/NuSea \cite{e866},
and HERMES \cite{hermesud} collaborations .
In particular, the E866/NuSea experimental results played an important
role in establishing the flavor asymmetry by clarifying
the $x$ dependence of $\bar u/ \bar d$. This new experimental
finding was a good opportunity of investigating a mysterious 
nonperturbative aspect of hadron structure.

Various models have been proposed to explain the unpolarized flavor
asymmetry. So far, meson-cloud type models are successful in explaining
the experimental results. For the explanation of these models
and other ideas, the authors suggest reading
the summary papers in Ref. \cite{udsum}.
Since most theoretical papers are written after the NMC finding,
the actual test of the proposed models should be done by predicting
unobserved quantities. In this sense, the polarized light-antiquark
flavor asymmetry should be an appropriate one.
In fact, there are already several papers on this topic
by phenomenological hadron models in Refs. \cite{models,fs,cs}.
It is particularly interesting to find that a meson-cloud model
and a chiral-soliton model predict totally different contributions 
to $\Delta \bar u - \Delta \bar d$, although their results are
similar in the unpolarized distribution $\bar u - \bar d$.

The situation of the polarized antiquark distributions is not as good
as the unpolarized one in the sense that the polarized whole sea-quark
distribution itself is not well determined at this stage.
Most parametrizations assume flavor symmetric polarized
antiquark distributions, which are then determined mainly
by the $g_1$ measurements.
As a result, there is much uncertainty in the antiquark
distributions at small and large $x$ \cite{aac,para}.
Although there are some semi-inclusive data \cite{semi} which could
be sensitive to the light antiquark flavor asymmetry, 
they are not accurate enough to provide strong constraint for
the polarized antiquark flavor asymmetry \cite{para,semi}.
However, the Relativistic Heavy Ion Collider (RHIC) \cite{rhic} and 
the Common Muon and Proton Apparatus for Structure and Spectroscopy
(COMPASS) \cite{compass} experiments should clarify the details
of the polarized antiquark distributions in a few years.
It is the right time to investigate
the antiquark flavor asymmetry $\Delta \bar u/\Delta \bar d$ by
possible theoretical models and to summarize various predictions.

In this paper, we intend to shed light on the virtual meson model
which has been successful in the unpolarized studies \cite{meson}.
The purpose of this paper is to extend the studies of the virtual
$\rho$-meson contributions by Fries and Sch\"afer in Ref. \cite{fs}.
In particular, we point out that the $g_2$ part of the polarized $\rho$
contributes to the polarized flavor asymmetry in addition to the
ordinary longitudinal part, which was calculated in Ref. \cite{fs}.
Because we show new $g_2$ terms in this paper and because the situation
is still confusing in the sense that another $\rho$-meson paper \cite{cs}
claims major differences from Ref. \cite{fs} in supposedly the same 
$\rho$ contributions, the detailed formalism is shown
in the following sections. 
The meson model was extended recently to a different direction
in Ref. \cite{cs} by including $\pi-\rho$ interference terms; however, 
this paper is intended to investigate a different kinematical aspect
within the meson model. 

The paper consists of the following. The formalism of $\rho$ contributions
to $\Delta \bar u - \Delta \bar d$ is presented in Sec. \ref{vector}.
Meson momentum distributions are obtained in Sec. \ref{meson},
and numerical results of $\Delta \bar u - \Delta \bar d$
are shown in Sec. \ref{results}. 
Our studies are summarized in Sec. \ref{concl}.

\section{Vector-meson contributions}
\label{vector}
\setcounter{equation}{0}

The cross section of polarized electron-nucleon scattering is generally
written in terms of lepton and hadron tensors:
\begin{equation}
\frac{d \sigma}{dE_e^\prime d\Omega_e^\prime} =
            \frac{|\vec p_e^{\, \prime}|}{| \vec p_e |} 
            \, \frac{\alpha^2}{(q^2)^2}
            \, L^{\mu \nu} (p_e, s_e, q)
            \, W_{\mu \nu} (p_N, s_N, q) ,
\label{eqn:cross1}
\end{equation}
where $\alpha$ is the fine structure constant,
$E_e^\prime$ and $\Omega_e^\prime$ are the scattered electron energy
and solid angle, and
$p_e$, $p_e^\prime$, $p_N$, and $q$ are initial electron, final electron,
nucleon, and virtual photon momenta, respectively.
The electron and nucleon spins are expressed by $s_e$ and $s_N$ 
with the normalization $s_e^2=s_N^2=-1$.
Throughout this paper, the convention $-g_{00}=g_{11}=g_{22}=g_{33}=+1$
is used so as to have, for example,
$p_N^2=(p_N^0)^2-\vec p_N^{\, \, 2}=m_N^2$.
Furthermore, the Dirac spinor is normalized as $u^\dagger u = E_e/m_e$
or $E_N/m_N$, where $E_e$ and $E_N$ are electron and nucleon energies,
and $m_e$ and $m_N$ are their masses.
The polarized lepton and hadron tensors are given by
\begin{align}
& L^{\mu\nu} (p_e, s_e, q) = 2 \, \big [  \, p_e^\mu {p_e^\prime}^\nu 
                             + p_e^\nu {p_e^\prime}^\mu
                    - (p_e \cdot p_e^\prime - m_e^2) \, g^{\mu\nu}
\nonumber \\
           & \ \ \ \ \ \ \ \ \ \ \ \ \ \ \ \ \ \ \ \ \ \ \ \ \ 
        - i \, \varepsilon^{\mu\nu\rho\sigma}
                       m_e q_\rho s_{e \, \sigma} \, \big ] ,
\\
& W_{\mu\nu} (p_N, s_N, q)  =  \frac{1}{2 \pi} 
                   \sum_X  \, (2\pi)^4 \, \delta^4(p_N+q-p_{_X})
\nonumber \\
           & \ \ \ \ \  \times         
                   <p_N,s_N |J_\mu (0)| \, X> \, <X|J_\nu (0)|\, p_N,s_N> ,
\end{align}
where the factor
$\varepsilon^{\mu\nu\rho\sigma}$ is the antisymmetric tensor
with the convention $\epsilon_{0123}=+1$.

\vspace{0.3cm}
\begin{figure}[h!]
\begin{center}
\includegraphics[width=0.35\textwidth]{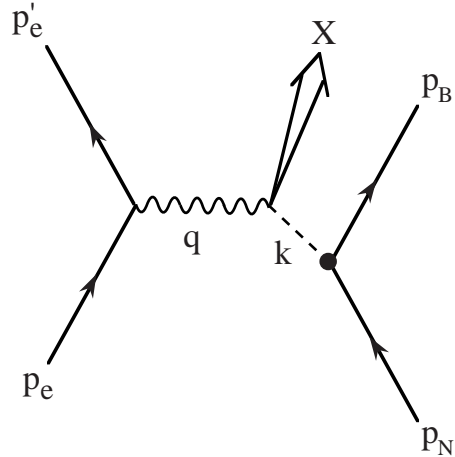}
\end{center}
\vspace{-0.5cm}
\caption{Virtual vector-meson contribution.}
\label{fig:vnb}
\end{figure}
\vspace{0.3cm}

Next, we consider the process in Fig. \ref{fig:vnb}, where
the nucleon splits into a virtual vector meson and a baryon,
then the virtual photon interacts with the polarized meson.
Because scalar mesons do not contribute directly to the polarized
structure functions due to spinless nature, the lightest vector meson,
namely $\rho$, is taken into account in this paper. In future, we may
extend the present studies by including heavier vector mesons.
As the final state baryon, the nucleon and $\Delta$ are considered.
Expressing the $VNB$ vertex multiplied by the meson propagator
as $J_{VNB}(k, s_V, p_N, s_N, p_B, s_B)$ and calculating
the cross section due to the process in Fig. \ref{fig:vnb}, we obtain
\begin{align}
& \frac{d \sigma}{dE_e^\prime d\Omega_e^\prime} = 
            \frac{|\vec p_e^{\, \prime}|}{| \vec p_e |} 
            \, \frac{\alpha^2}{(q^2)^2}
            \, L^{\mu \nu} (p_e, s_e, q)
\nonumber \\
& \times \, \int \frac{d^3 p_B}{(2\pi)^3} \, \frac{m_B}{2 \pi E_B}
        \sum_{X, \lambda_V, \lambda_B} | J_{VNB} |^2
          <k,s_V |J_\mu (0)| \, X> 
\nonumber \\ 
&  \times \, <X|J_\nu (0)|\, k,s_V>  
           (2 \pi)^4 \, \delta^4(k+q-p_{_X})  .
\end{align}
Here, $k$ and $s_V$ indicate the meson momentum and spin.
This equation has the same form as Eq. (\ref{eqn:cross1}).
Therefore, the last part is identified with a vector-meson
contribution to the nucleon tensor:
\begin{align}
W_{\mu\nu} (p_N, s_N, & q)  =  \int \frac{d^3 p_B}{(2\pi)^3}
            \, \frac{2 m_V m_B}{E_B} 
\nonumber \\
& \times   \sum_{\lambda_V,\lambda_B} | J_{VNB} |^2
            \, W_{\mu\nu}^{(V)} (k, s_V, q)  ,
\label{eqn:w-meson}
\end{align}
where $m_V$ is the meson mass, and the meson tensor is defined by
\begin{align}
W_{\mu\nu}^{(V)} & (k, s_V, q)  = \frac{1}{4 \pi m_V} \sum_{X}
          (2 \pi)^4 \, \delta^4(k+q-p_{_X})
\nonumber \\
           &  \times         
         <k,s_V |J_\mu (0)| \, X> \, <X|J_\nu (0)|\, k,s_V> .
\end{align}

In this way, the vector-meson contribution to the nucleon tensor
is expressed in terms of the $VNB$ vertex and the meson tensor. 
Because we are interested in meson effects on the polarized parton
distributions in the nucleon, we try to project the $g_1$ part out
from the nucleon tensor.
The definition of the $g_1$ and $g_2$ structure functions is given
in the asymmetric part of the nucleon tensor:
\begin{align}
W^A_{\mu \nu} (p_N, &  s_N, q) = 
       i \, \varepsilon_{\mu\nu\rho\sigma} \, q^{\, \rho}
\bigg [ \,  s_N^{\, \sigma} \, \frac{g_1}{p_N \cdot q} 
\nonumber \\
&        + ( p_N \cdot q \, s_N^{\, \sigma} - s_N \cdot q \, p_N^{\, \sigma} ) 
          \, \frac {g_2}{(p_N \cdot q)^2} \,
\bigg ] .
\end{align}
In order to discuss each structure function separately, 
a projection operator
\begin{equation}
P^{\mu \nu} = - \frac{m_N^2}{2 \, p_N \cdot q} 
         \, i \, \varepsilon^{\mu\nu\alpha\beta} \, q_\alpha \, s_{N \, \beta} ,
\end{equation}
is then applied to give 
\begin{equation}
P^{\mu \nu} \, W^A_{\mu \nu} (p_N, s_N, q) = 
     \frac{m_N^2}{(p_N \cdot q)^2} \left [ q^2 + (s_N \cdot q)^2 \right ] \, g_1
        - \gamma^2 \, g_2 .
\label{eqn:pw}
\end{equation}
Here, $\gamma$ is defined by
\begin{equation}
\gamma^2 = \frac{4 \, x^2 \, m_N^2}{Q^2} ,
\end{equation}
with $Q^2=-q^2$.
In the same way, $g_1$ and $g_2$ structure functions of the vector
meson are defined in the asymmetric part of the tensor.
Operating the projection also on the meson tensor, we obtain
\begin{equation}
P^{\mu \nu} W^{(V)A}_{\mu \nu} (k, s_V, q)  =  \frac{m_N}{m_V}  \,
                 [ \,  A_1  \, g_1^{V} (k,q)
                     + A_2  \, g_2^{V} (k,q) \, ] ,
\label{eqn:pc}
\end{equation}
where $A_1$ and $A_2$ are given by
\begin{align}
A_1 & = \frac{m_N \, m_V}{p_N \cdot q \, k \cdot q} \,
          ( s_N \cdot q  \, s_V \cdot q - q^2 s_N \cdot s_V ) ,
\\
A_2 & =  \frac{m_N \, m_V \, q^2}{p_N \cdot q \, (k \cdot q)^2} \,
           ( s_N \cdot k \, s_V \cdot q  - k \cdot q \, s_N \cdot s_V ) .
\end{align}

From Eqs. (\ref{eqn:w-meson}), (\ref{eqn:pw}), and (\ref{eqn:pc}),
the meson contribution to the nucleon structure functions becomes
\begin{align}
& \frac{m_N^2}{(p_N \cdot q)^2} \,
      [ \, q^2 + (s_N \cdot q)^2 \, ] \, g_1 (p_N, q) 
     - \gamma^2 \, g_2 (p_N, q) 
\nonumber \\ 
& \ \ \ \ =  \int \frac{d^3 p_B}{(2\pi)^3}
            \, \frac{2 m_N m_B}{E_B} 
  \sum_{\lambda_V, \lambda_B} | J_{VNB} |^2
\nonumber \\ 
& \ \ \ \ \ \ \ \ \ \ \ \ \times
         \,  [ \,  A_1 \, g_1^{V} (k,q)
                 + A_2 \, g_2^{V} (k,q) \, ]  .
\end{align}
Then, the above integration variables ($p_B^x$, $p_B^y$, $p_B^z$)
are changed for the meson momentum fraction $y$,
the transverse momentum $k_\perp$ of the meson,
and the angle $\phi$ between $\vec k_\perp$
and the transverse spin vector of the nucleon ($\vec s_N^{\, T}$):
\begin{equation}
y = \frac{k \cdot q}{p_N \cdot q}\ , \ \ \ \ \ 
\vec k \cdot \vec s_N^{\, T} = k_\perp  \tau_N  \cos \phi ,
\end{equation}
with $\tau_N = | \vec s_N^{\, T} |$.
Then, the meson contribution is expressed as
\begin{align}
& \frac{m_N^2}{(p_N \cdot q)^2} \,
      [ \, q^2 + (s_N \cdot q)^2 \, ] \, g_1 (x, Q^2) 
     - \gamma^2 \, g_2 (x, Q^2) 
\nonumber \\ 
& =  \int_x^1 \frac{dy}{y}
         \,  [ \,  B_1 (y) \, g_1^{V} (x/y, Q^2)
                 + B_2 (y) \, g_2^{V} (x/y, Q^2) \, ] .
\label{eqn:g1g2v}
\end{align}
The upper limit of the $y$-integration range is taken as 1
by considering the vector-meson mass smaller than
the nucleon mass. However, one should be careful in extending
the present studies to other mesons with larger masses.
The meson momentum distributions are expressed as
\begin{align}
B_{1,2} (y)  = & \int_0^{(\vec k_\perp^{\, 2})_{max}}  
                d \vec k_\perp^{\, 2}
                 \int_0^{2\pi} d\phi 
                 \frac{| \vec p_N| \, m_N \, m_B}{(2\pi)^3 E_B} \,
                 \frac{\partial y'}{\partial y}  \, y
\nonumber \\
         & \times  \sum_{\lambda_V, \lambda_B}
                 | J_{VNB} |^2  \, A_{1,2} \ ,
\label{eqn:b12}
\end{align}
where $y'$ is the longitudinal momentum fraction defined in the
meson momentum
\begin{equation}
\vec k = \vec k_\perp + y' \vec p_N .
\end{equation}
In the infinite momentum frame $|\vec p_N| \rightarrow \infty$,
$y$ and $y'$ are related by
\begin{equation}
y' =  \frac{y}{1+\sqrt{1+\gamma^2}}
     \,  \bigg [ \, 1 + \sqrt{ 1+ \frac{\gamma^2}{y^2 m_N^2}
                      ( \vec k_\perp^2 + m_V^2 ) } \,   \bigg ] .
\end{equation}
Because time-ordered perturbation theory is used for the
reaction in Fig. \ref{fig:vnb} as explained in Sec. \ref{meson},
the vector meson is taken as an on-shell particle:
$k^2 = m_V^2$ in the above derivation.
The partial derivative $\partial y' /\partial y$
can be calculated from this expression.
In the infinite momentum frame, the momentum fraction $y'$ 
has to satisfy the kinematical condition $0 \le y' \le 1$,
namely the meson $V$ and the baryon $B$ should move
in the forward direction. 
The maximum transverse momentum is given by 
\begin{equation}
(\vec k_\perp^{\, 2})_{max} =
             \frac{m_N^2}{\gamma^2} \, (\sqrt{1+\gamma^2}+1) \,
                 (\sqrt{1+\gamma^2}+1-2y)
           - m_V^2 .
\end{equation}
Practically, it does not matter to take the upper bound
$(\vec k_\perp^{\, 2})_{max} \rightarrow \infty$ in Eq. (\ref{eqn:b12})
at small-$x$ where the antiquark distributions play a major role,
because
$(\vec k_\perp^{\, 2})_{max}$ $(\sim Q^2 (1-y)/x^2 \gg m_N^2 )$
is beyond the vertex momentum cutoff region discussed in
Sec. \ref{meson}. The contribution to the integral
between $(\vec k_\perp^{\, 2})_{max}$ and $\vec k_\perp^{\, 2}=\infty$
is extremely small in general.
Furthermore, the upper bound becomes 
$(\vec k_\perp^{\, 2})_{max} \rightarrow \infty$
in the limit $Q^2 \rightarrow \infty$, and it  
is consistent with the previous publications \cite{fs,cs}.
In this way, the meson contribution is expressed in terms of 
the meson structure functions convoluted with the meson
momentum distributions in the nucleon.

Using the integration variables $y$, $\vec k_\perp ^2$, and $\phi$,
we express the coefficients $A_1$ and $A_2$ as 
\begin{align}
A_1 = & \lambda_N \lambda_V \bigg [ 1 + \frac{\vec k_\perp ^2}{y y' \, m_N^2}
                                        ( \sqrt{1+\gamma^2} - 1 ) \bigg ] 
\nonumber \\
 & \ \ \ \  - \gamma^2 \tau_N \lambda_V  \cos \phi
               \, \frac{k_\perp}{y \, m_N} ,
\nonumber \\
A_2 = & \frac{\gamma^2 m_V^2}{y^2 m_N^2} \bigg [  \, - \lambda_N \lambda_V 
    + \tau_N \lambda_V \! \cos \phi \, \frac{ k_\perp}{y' \, m_N} \,
       ( \sqrt{1+\gamma^2} - 1 ) \, \bigg ]  ,
\end{align}
in the limit $|\vec p_N| \rightarrow \infty$.
Detailed calculations indicate that $\phi$ dependence can be extracted
out from another part of the integrand in Eq.(\ref{eqn:b12}) as
\begin{equation}
\frac{| \vec p_N| \, m_N \, m_B}{(2\pi)^3 E_B} \,
   \frac{\partial y'}{\partial y}  \, y \sum_{\lambda_B} | J_{VNB} |^2  
\equiv C_L^{\lambda_V} + \tau_N \cos \phi \, C_T^{\lambda_V}  \ .
\label{eqn:clt}
\end{equation}
Then, after the $\phi$ integration, Eq. (\ref{eqn:b12}) becomes
\begin{align}
B_1 (y)  & = \sum_{\lambda_V}
             \lambda_V \, [ \lambda_N \, f_{1L}^{\lambda_V} (y)
                               -\tau_N^2 \, f_{1T}^{\lambda_V} (y) ] , 
\\
B_2 (y)  & = \sum_{\lambda_V}
             \lambda_V \, [ - \lambda_N \, f_{2L}^{\lambda_V} (y)
                               + \tau_N^2 \, f_{2T}^{\lambda_V} (y) ] ,
\end{align}
where $\lambda_V$ dependence is explicitly denoted in meson momentum
distributions, which are defined by
\begin{align}
f_{1L}^{\lambda_V} (y)  & =
 \int_0^{(\vec k_\perp^{\, 2})_{max}}  \! \! \! d \vec k_\perp^{\, 2}
     2 \pi \, C_L^{\lambda_V} \,
     \bigg [ 1 + \frac{\vec k_\perp ^2}{yy' m_N^2} 
                  ( \sqrt{1+\gamma^2} -1 ) \bigg ] \, ,
\label{eqn:f1l}
\\
f_{1T}^{\lambda_V} (y)  & = 
 \int_0^{(\vec k_\perp^{\, 2})_{max}}  \! \! \! d \vec k_\perp^{\, 2}
    \gamma^2 \, \pi \, C_T^{\lambda_V} \, \frac{k_\perp}{y m_N} \, ,
\label{eqn:f1t}
\\
f_{2L}^{\lambda_V} (y)  & = 
 \int_0^{(\vec k_\perp^{\, 2})_{max}}  \! \! \! d \vec k_\perp^{\, 2}
\gamma^2 \, 2 \pi \, C_L^{\lambda_V} \, \frac{m_V^2}{y^2 m_N^2}  \, ,
\label{eqn:f2l}
\end{align}
\begin{align}
f_{2T}^{\lambda_V} (y)  & =
 \int_0^{(\vec k_\perp^{\, 2})_{max}}  \! \! \! d \vec k_\perp^{\, 2}
 \gamma^2 \, \pi \, C_T^{\lambda_V} \, \frac{k_\perp m_V^2}{y' y^2 m_N^3} 
                  ( \sqrt{1+\gamma^2} -1 )         \, .
\label{eqn:f2t}
\end{align}
Because the functions $f_{1T}^{\lambda_V}$, $f_{2L}^{\lambda_V}$, and
$f_{2T}^{\lambda_V}$ are proportional to $\gamma^2$, they vanish
in the limit $Q^2 \rightarrow \infty$.
As it is obvious from Eq. (\ref{eqn:g1g2v}),
it is necessary to consider both longitudinal and transverse 
polarizations for the nucleon in order to extract the $g_1$ part.
In addition, the $g_2$ structure function of the meson contributes.
The function $f_{1L}^{\lambda_V} (y)$ is the ordinary meson momentum
distribution with the momentum fraction $y$ in the longitudinally
polarized nucleon. The function $f_{1T}^{\lambda_V} (y)$ is
the distribution in the transversely polarized nucleon.
On the other hand, $f_{2L}^{\lambda_V} (y)$ and $f_{2T}^{\lambda_V} (y)$
are the distributions associated with $g_2$ of the vector meson.
Expressing Eq. (\ref{eqn:g1g2v}) in terms of the nucleon and meson
helicities, $\lambda_N$ and $\lambda_V$, we obtain
\begin{align}
& (\lambda_N^2 - \tau_N^2 \, \gamma^2 ) \, g_1 (x, Q^2) 
     - \gamma^2 \, g_2 (x, Q^2) 
\nonumber \\ 
& = \sum_{\lambda_V}  \lambda_V \int_x^1 \frac{dy}{y}
    \, \big [ \, \big \{ \lambda_N \, f_{1L} (y)
                                - \tau_N^2 \, f_{1T} (y) \big \}
                               \, g_1^{V} (x/y, Q^2)
\nonumber \\
& \ \ \ \  + \big \{ - \lambda_N \, f_{2L} (y)
                                + \tau_N^2 \, f_{2T} (y) \big \}
                               \, g_2^{V} (x/y, Q^2)        \, \big ] .
\label{eqn:g1g2v2}
\end{align}

Combining the longitudinal polarization $\lambda_N =1$ ($\tau_N=0$)
with the transverse polarization $\tau_N=1$ ($\lambda_N = 0$),
we can extract the $g_1$ part as
\begin{align}
& g_1 (x, Q^2) = \frac{1}{1+\gamma^2} \int_x^1 \frac{dy}{y}
\nonumber \\ 
& \ \ \times
     \, \big [ \, \big \{ \Delta f_{1L} (y) + \Delta f_{1T} (y) \big \} 
     \, g_1^{V} (x/y, Q^2)
\nonumber \\
& \ \ \ \ \ 
                - \big \{ \Delta f_{2L} (y) + \Delta f_{2T} (y) \big \} 
     \, g_2^{V} (x/y, Q^2)
  \, \big ]   ,
\label{eqn:g1v}
\end{align}
where the functions $\Delta f_i^{VN}(y)$ with $i$=$1L$, $2L$, $1T$,
and $2T$ are defined by 
\begin{equation}
\Delta f_i (y) = f_i^{\lambda_V=+1} (y)
               - f_i^{\lambda_V=-1} (y)  .
\label{eqn:dfy}
\end{equation}
The $g_2$ part is obtained in the same way as
\begin{align}
& g_2 (x, Q^2) = \frac{1}{1+\gamma^2} \int_x^1 \frac{dy}{y}
\nonumber \\ 
& \ \ \times
     \, \big [ \, \big \{ - \Delta f_{1L} (y) 
                   + \Delta f_{1T} (y) / \gamma^2 \big \}
     \, g_1^{V} (x/y, Q^2)
\nonumber \\
& \ 
           +  \big \{ \Delta f_{2L} (y) 
                - \Delta f_{2T} (y) / \gamma^2  \big \} 
     \, g_2^{V} (x/y, Q^2)
  \, \big ]  .
\label{eqn:g2v}
\end{align}
In the limit $Q^2 \rightarrow \infty$, namely $\gamma^2 \rightarrow 0$,
only the momentum distribution $\Delta f_{1L} (y)$ remains finite,
and Eq. (\ref{eqn:g1v}) agrees with the expression in Ref. \cite{fs}.

In Eq. (\ref{eqn:g1v}), there are additional terms associated with
$g_2$ of the meson. For discussing  these $g_2$ type contributions
to $\Delta \bar u - \Delta \bar d$, $g_2^V$ is approximated
by the Wandzura-Wilczek (WW) relation \cite{ww} by neglecting
higher-twist terms:
\begin{equation}
g_2^{V (WW)} (x,Q^2) 
             =  - g_1^V(x,Q^2) + \int_x^1 \frac{dy}{y} g_1^V(y,Q^2) .
\end{equation}
Then, providing the leading-order expression for $g_1^V$, we have
\begin{align}
g_2^{V (WW)} (x,Q^2) =
  \frac{1}{2} \sum_i \, e_i^2 \, [ \, & \Delta q_i^{V (WW)} (x,Q^2) 
\nonumber \\
                          + & \Delta \bar q_i^{V (WW)} (x,Q^2) \, ] .
\label{eqn:ww}
\end{align}
The above WW distributions are defined by
\begin{equation}
\Delta \bar q_i^{V (WW)}  (x,Q^2) 
            =  -  \Delta \bar q_i^V (x,Q^2) 
               + \int_x^1 \frac{dy}{y} 
                   \,  \Delta \bar q_i^V (y,Q^2) ,
\end{equation}
and the same equation for $\Delta q_i^{V (WW)}  (x,Q^2)$.
From these equations, we obtain a vector meson contribution
to the polarized antiquark distribution $\Delta \bar q_i$ in the proton as
\begin{align}
& \Delta \bar q_i^{VNB} (x, Q^2) = \frac{1}{1+\gamma^2} \int_x^1 \frac{dy}{y}
\nonumber \\ 
& \ \times
     \, \big [ \, \big \{ \Delta f_{1L} (y) + \Delta f_{1T} (y) \big \} 
     \, \Delta \bar q_i^V (x/y, Q^2)
\nonumber \\
& \ \ \ \ 
                - \big \{ \Delta f_{2L} (y) + \Delta f_{2T} (y) \big \} 
     \, \Delta \bar q_i^{V (WW)} (x/y, Q^2)
  \, \big ]   .
\label{eqn:dPi}
\end{align}
If this kind of vector-meson contribution is the only source for
the polarized flavor asymmetry, the $\Delta \bar u - \Delta \bar d$
distribution is then calculated by taking the difference 
$\Delta \bar q_ {i=u}^{VNB} - \Delta \bar q_ {i=d}^{VNB}$
in the above equation.

\section{Meson momentum distributions}
\label{meson}
\setcounter{equation}{0}

In order to estimate the meson contributions numerically, it is necessary
to calculate the momentum distributions $\Delta f_i^{VN}(y)$ of the meson.
We calculate them by considering the vector-meson creation processes 
$N \rightarrow V N'$ and $N \rightarrow V \Delta$ through 
the interactions
\begin{align}
V_{VNN} & =  \widetilde \phi_V^* \cdot \widetilde T \, 
             F_{VNN} (k) \, \,
             \overline u (p_N', s_N') \, \epsi^{\mu \, *}
\nonumber \\
& \ \ \ \ 
      \times  \bigg[ \, g_V \gamma_\mu 
         - \frac{f_V}{2 m_N} \, i \, \sigma_{\mu\nu} 
                                \khat^\nu  \, \bigg ] \,   
          \, u (p_N, s_N) ,
\label{eqn:vvnn}
\\
V_{V N \Delta} & = \widetilde \phi_V^* \cdot \widetilde T \, 
             F_{VN \Delta} (k) \, \,
             \overline U_\nu (p_\Delta, s_\Delta) 
             \, \frac{f_{VN\Delta}}{m_V} \, \gamma_5 \, \gamma_\mu 
\nonumber \\
&  \times
             \,  \big [ \, \khat^\mu \, \epsi^{\nu \, *}
             - \khat^\nu \, \epsi^{\mu \, *} \, \big ]
              \,  u (p_N, s_N) ,
\label{eqn:vvnd}
\end{align}
where $u (p_N, s_N)$ is the Dirac spinor, 
$U^\mu (p_\Delta, s_\Delta)$ is the Rarita-Schwinger spinor,
and $\epsi^\mu$ is the polarization vector of the vector meson. 
The $VNN$ and $VN \Delta$ coupling constants are denoted
as $g_V$, $f_V$, and $f_{VN\Delta}$, and form factors are denoted as
$F_{VNN} (k)$ and $F_{VN \Delta} (k)$.
Isospin dependence is taken into account by the factor
$\widetilde \phi_V^* \cdot \widetilde T$, and it is defined
in terms of a reduced matrix element and a Clebsch-Gordan coefficient
\cite{skflux,edmonds}
\begin{align}
< \! B \, | \, \widetilde \phi_V^* \cdot \widetilde T \, | \, N \! > &
=  (-1)^{M_V} \frac{<T_B \parallel 
             \widehat T \parallel T_N \! >}{\sqrt{2T_B+1}}
\nonumber \\
 &  \times  < T_N M_N : 1 - M_V | T_B M_B \! > 
\label{eqn:isospin}
\end{align}
with $<1/2 \parallel \widehat T \parallel \! 1/2  > = \sqrt{6}$ 
and  $<3/2 \parallel \widehat T \parallel \! 1/2  > = 2$.
Here, $T_N$ and $T_B$ denote isospins of the nucleon and the baryon, 
respectively, and $M_N$ and $M_B$ are their third components.

From these vertices, the meson momentum distribution $f_M (y)$ can be
calculated together with the baryon distribution $f_B (y)$. They are
supposed to satisfy the relation $f_M (y)=f_B (1-y)$ because of
charge and momentum conservations. However, it is known
that the covariant calculation could violate this relation because
a derivative coupling introduces off-shell dependence. 
A possible solution \cite{julich,mt} is to use the time-ordered
perturbation theory (TOPT), instead of the covariant formalism.
Although the four-momentum conservation is satisfied 
at the $VNB$ vertex in the covariant formalism,
the energy is not conserved in the TOPT \cite{topt}.
If there is no off-shell dependence at a vertex,
the TOPT agrees certainly with the ordinary covariant theory 
by collecting all the time-ordered diagrams.
However, the off-shell dependence due to the derivative coupling
complicates the problem. It leads to a freedom in defining
the vertex momentum $\widehat K$ in Eqs. (\ref{eqn:vvnn}) and
(\ref{eqn:vvnd}). The following two possibilities are considered
in Refs. \cite{fs,julich}: 
\begin{align}
\text{(A)}  \ \ \ & \khat = k  = (E_V, \vec k),
\nonumber \\
\text{(B)}  \ \ \ & \khat = \kd = p_N - p_B = (E_N - E_B, \vec k),
\end{align}
where $E_V=\sqrt{m_V^2 + \vec k^{\, 2}}$.
There is another off-shell dependence from the vertex form factors,
and it is discussed in Sec. \ref{results}.

From the $VNN$ interaction vertex in Eq. (\ref{eqn:vvnn}),
we obtain
\begin{align}
&  \sum_{\lambda_N'}  |V_{VNN}|^2 = 
   | \widetilde \phi_V^* \cdot \widetilde T |^2 \,  
   F_{VNN}^{\, 2} (k) \, \frac{1}{4 \, m_N^2} \,
   \bigg [ \, g_V^2 \, \big \{ - \kd^2 
\nonumber \\
& \ \ \ 
+ 2 \, (  \pn \cdot \epsi \, \pnd \cdot \epsi^*  
        + \pn \cdot \epsi^*  \pnd \cdot \epsi   ) 
+ 2 m_N m_V s_N \cdot \overline s_V  \big \}
\nonumber \\
&  + g_V \, f_V \, \big \{ - 2 \kd \cdot \khat
             + 2 m_N m_V s_N \cdot \hat s_V       
\nonumber \\
& \ \ \ \ \ \ \ \ \ \ \ \ \ \ 
        - 2 \pnd \cdot \khat \, s_N \cdot s' 
        -  \khat \cdot \epsi  \, \kd \cdot \epsi^*
                - \khat \cdot \epsi^* \kd \cdot \epsi \, \} 
\nonumber \\
& + \frac{f_V^2}{4 m_N^2} \, \big \{ \, 
    (\khat^2 + \khat \cdot \epsi \, \khat \cdot \epsi^* ) 
    ( \kd^2 - 4 m_N^2 )
     + 4 \pn \cdot \khat  \pnd \cdot \khat
\nonumber \\
& \ \ \  
     - 2 \khat^2 (  \pn \cdot \epsi  \, \pnd \cdot \epsi^* 
                  + \pn \cdot \epsi^*   \pnd \cdot \epsi   )
     + 4 m_N^2 \khat^2 s_N \cdot s'
\nonumber \\
& \ \ \ 
     - 4 m_N m_V \pnd \cdot \khat s_N \cdot \hat s_V 
     - 2 m_N m_V \khat^2 s_N \cdot \overline s_V
\nonumber \\
& \ \ \ 
     + 2 ( \pn \cdot \khat + \pnd \cdot \khat )
            (  \pn \cdot \epsi   \, \khat \cdot \epsi^*
             + \pn \cdot \epsi^*    \khat \cdot \epsi  )
\nonumber \\
& \ \ \ 
     - 2 \pn \cdot \khat (  \khat \cdot \epsi  \, \kd \cdot \epsi^*
                           +\khat \cdot \epsi^*   \kd \cdot \epsi   )
  \, \big \} \, \bigg ] \ ,
\label{eqn:vnn}
\end{align}
where ${s'} ^\mu$, $\hat s_V ^\mu$, $\overline s_V ^\mu$
are defined by
\begin{align}
{s'} ^{\, \mu}   &  = - \frac{i}{m_N} \, \epsi^{\mu\nu\alpha\beta} 
               \, \epsi_\nu^* \, \epsi_\alpha \, p_{N \beta}  ,
\\
\hat s_V ^{\, \mu} &  = - \frac{i}{m_V} \, \epsi^{\mu\nu\alpha\beta} 
                \, \epsi_\nu^* \, \epsi_\alpha \, \widehat K_\beta ,
\\
\overline s_V ^{\, \mu} &  = - \frac{i}{m_V} \, \epsi^{\mu\nu\alpha\beta} 
                \, \epsi_\nu^* \, \epsi_\alpha \, \overline k_\beta .
\end{align}
The meson polarization vector is given by
\begin{equation}
\epsi^\mu  = \bigg ( \frac{\vec k \cdot \vec\epsi_{\lambda_V}}{m_V},
           \vec\epsi_{\lambda_V} 
          + \frac{\vec k \cdot \vec\epsi_{\lambda_V}}{m_V (E_V + m_V)} 
               \vec k \bigg )  ,
\end{equation}
where $\lambda_V$ is the meson helicity.
The spherical unit vector $\vec\epsi_{\lambda_V}$ is defined
in the frame with the $\hat z'$ axis parallel to $\vec k$:
\begin{equation}
\vec\epsi_{\pm 1} = \mp \frac{1}{\sqrt{2}} ( \hat x ' \pm i \hat y' ) ,
\ \ \ 
\vec\epsi_{0} = \hat z' .
\end{equation}

In the same way, the $VN\Delta$ term is calculated from
Eq. (\ref{eqn:vvnd}) as
\begin{align}
& \sum_{\lambda_\Delta} |V_{VN\Delta}|^2 =
   - | \widetilde \phi_V^* \cdot \widetilde T |^2 \,  
                       F_{VN \Delta}^{\, 2} (k) \,
\frac{f_{VN\Delta}^2}{6 \, \mn \md^3  \mv^2}
\nonumber \\
& \times  \bigg [ \,
- 2 \md^2 p_N \cdot \khat p_\Delta \cdot \khat 
+ 2 \mn \md^3 (\khat^2 + \khat \cdot \epsi \khat \cdot \epsi^*)
\nonumber \\
& \ \ 
+ 2 \pn \cdot \pd \big \{ \md^2 \khat^2 
                         + \md^2 \khat \cdot \epsi \khat \cdot \epsi^* 
                         + \khat^2 \pd \cdot \epsi \, \pd \cdot \epsi^* 
\nonumber \\
&  \ \ \ \ \ \ \ \ 
- \pd \cdot \khat ( \pd \cdot \epsi \khat \cdot \epsi^* 
                   +\pd \cdot \epsi^* \khat \cdot \epsi )
- (\pd \cdot \khat)^2 \big \}
\nonumber \\
& \ \ 
+ \md^2 \big \{ - \pn \cdot \khat ( \pd \cdot \epsi \khat \cdot \epsi^* 
                               +\pd \cdot \epsi^* \khat \cdot \epsi )
\nonumber \\
&   \ \ \ \ \ \ \ \ \ \ \ \ \ 
               - \pd \cdot \khat  ( \pn \cdot \epsi \khat \cdot \epsi^* 
                               +\pn \cdot \epsi^* \khat \cdot \epsi )
\nonumber \\
&  \ \ \ \ \ \ \ \ \ \ \ \ \ 
               + \khat^2  ( \pn \cdot \epsi \pd \cdot \epsi^* 
                            +\pn \cdot \epsi^* \pd \cdot \epsi ) \, \big \}
\nonumber \\
& \ \ 
+ \mn s_N \cdot s'  \big \{  2 \mn (\pd \cdot \khat)^2 
                      + \md^2 (\mn +\md) \khat^2   \big \}
\nonumber \\
& \ \ 
- \mn \mv s_N \cdot \overline s_V 
           \big \{ 2 (\pd \cdot \khat)^2 + \md^2 \khat^2   \big \}
\nonumber \\
& \ \ 
- 4 \mn \md^2 \mv \pd \cdot \khat  \,  s_N \cdot \hat s_V
\nonumber \\
& \ \ 
+ 2 \mn^2 \md \mv (\khat^2 s_N \cdot s_1 
                   - \pd \cdot \khat s_N \cdot \hat s_2) \, \bigg ] ,
\label{eqn:vnd}
\end{align}
where $s_1^\mu$ and $\hat s_2^\mu$ are defined by
\begin{align}
s_1 ^\mu  &  = \frac{i}{m_N^2 m_V} 
                \, \epsi^{\mu\nu\alpha\beta} 
                \, (  p_N \cdot \epsi \, \epsi_\nu^* 
                  - p_N \cdot \epsi^* \epsi_\nu )
                   \, p_{N \alpha} \, \overline k_\beta  \, , 
\\
\hat s_2 ^\mu  &  = \frac{i}{m_N \, m_\Delta \, m_V}
                \, \epsi^{\mu\nu\alpha\beta} 
                \, (  p_N \cdot \epsi \, \epsi_\nu^* 
                    - p_N \cdot \epsi^* \epsi_\nu ) 
                   \, p_{\Delta \alpha} \, \khat_\beta  \, .
\end{align}

In Sec. \ref{vector}, the term $J_{VNB}$  is defined by
the vertex $V_{VNB}$ multiplied by the meson propagator. 
The propagator is the addition of two time-ordered terms. However,
only the first one remains finite in an infinite momentum frame
$p_N\rightarrow \infty$:
\begin{align}
& \frac{1}{2 E_V (E_N-E_V-E_B)}
          + \frac{1}{2 E_V (E_B-E_V-E_N)} 
\nonumber \\
& \ \ \ \ \ \ \ \ \ \ 
       = \frac{1}{y' (m_N^2 -m_{VB}^2)} \ ,
\end{align}
where $m_{VB}^2$ is defined by
\begin{equation}
m_{VB}^2 = (k+p_B)^2 = \frac{m_V^2 + \vec k_\perp^{\, 2}}{y'}
                    +\frac{m_B^2 + \vec k_\perp^{\, 2}}{1-y'} .
\label{eqn:mvb2}
\end{equation}
Therefore, $J_{VNB}$ is expressed as
\begin{equation}
J_{VNB} = \frac{1}{y' (m_N^2 -m_{VB}^2)} \, V_{VNB} .
\label{eqn:jv}
\end{equation}
Using Eqs. (\ref{eqn:vnn}), (\ref{eqn:vnd}), and (\ref{eqn:jv})
together with Eqs. (\ref{eqn:clt}) and (\ref{eqn:f1l})--(\ref{eqn:f2t}), 
we can calculate the meson momentum distributions.
The actual calculations are partially done by a Maple program.
Obtained expressions are rather lengthy, so that the results
are written in Appendix. 

The momentum distributions are numerically calculated by using 
the expressions in Appendix. However, the derivation of these
analytical expressions is complicated,
and it could easily lead to a calculation mistake. 
In order to avoid this kind of failure, we calculated
the momentum distributions numerically in an independent way
directly from Eqs. (\ref{eqn:vnn}) and (\ref{eqn:vnd}), 
and we confirmed  that they indeed agree on
the results in Appendix.

\begin{figure}[t!]
\includegraphics[width=0.41\textwidth]{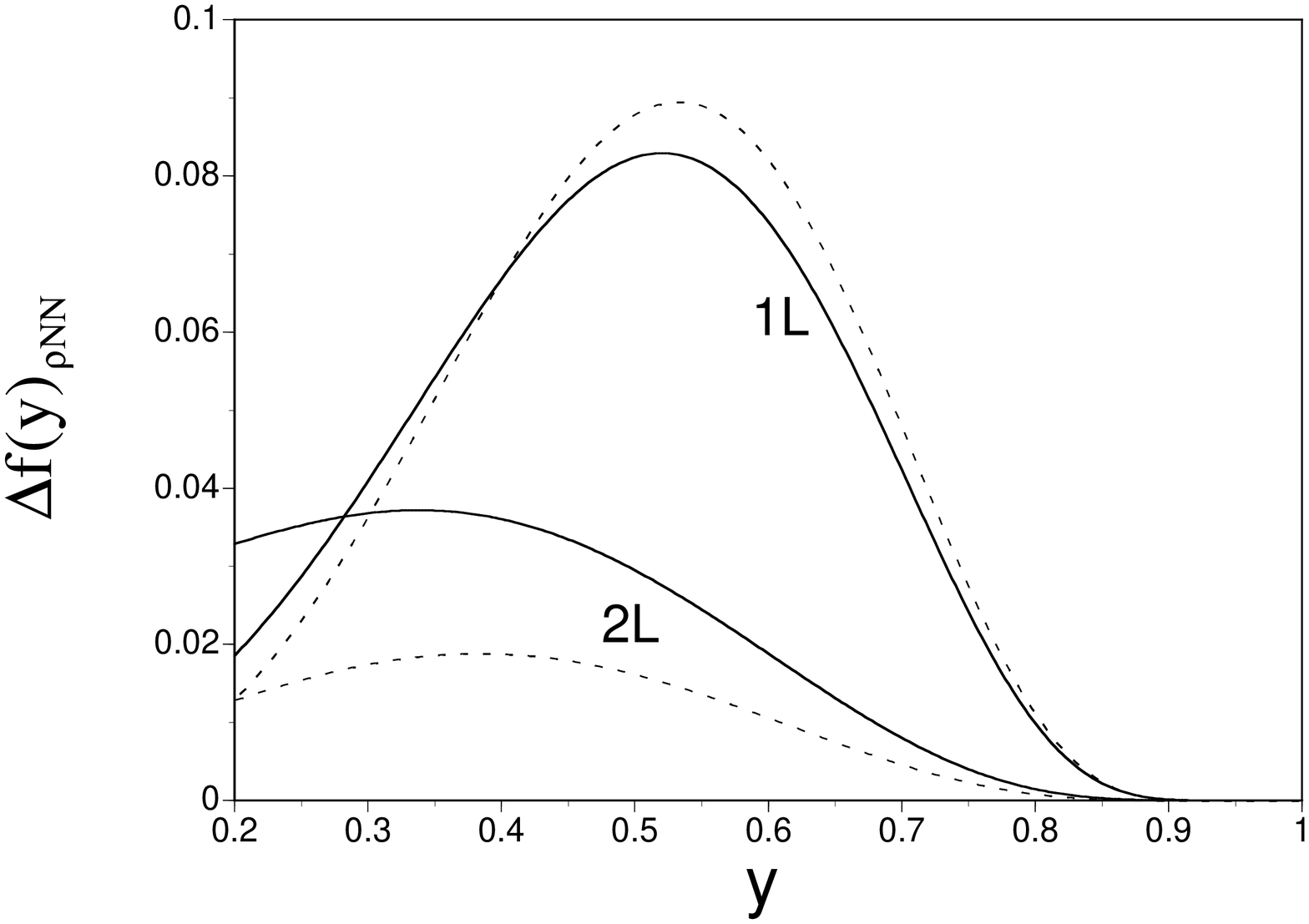}
\vspace{-0.8cm}
\caption{Meson momentum distributions $\Delta f_{1L}(y)$ 
         and $\Delta f_{2L}(y)$ from the $\rho NN$ process.
         The solid and dashed curves are obtained at
         $Q^2$=1 and 2 GeV$^2$, respectively, with $x$=0.2.
         The isospin factors are taken out from the distributions
         as explained in the text.}
\label{fig:dfbnl}
\vspace{0.7cm}
\includegraphics[width=0.41\textwidth]{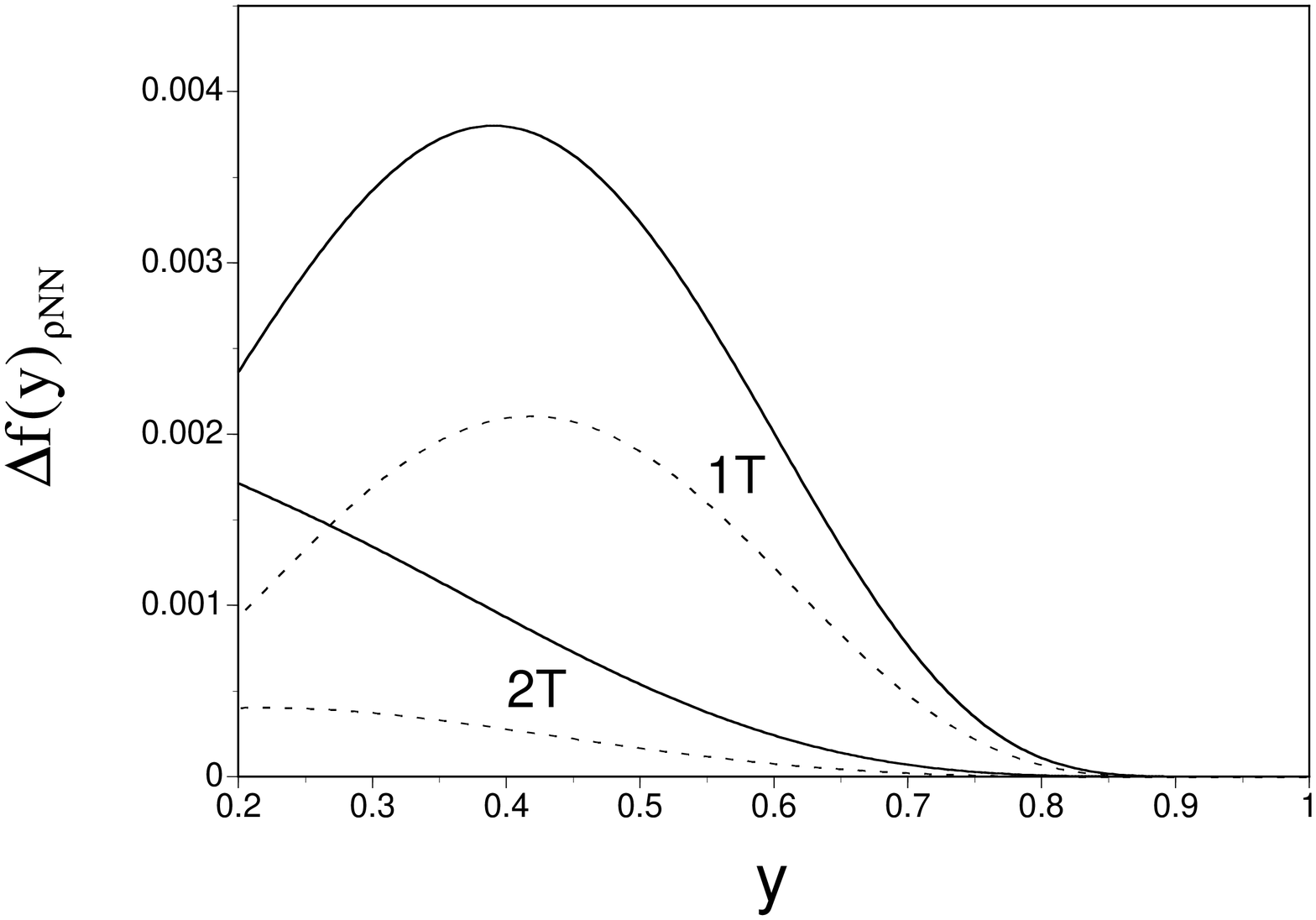}
\vspace{-0.8cm}
\caption{Meson momentum distributions $\Delta f_{1T}(y)$ 
         and $\Delta f_{2T}(y)$ from the $\rho NN$ process.
         The solid and dashed curves are obtained at
         $Q^2$=1 and 2 GeV$^2$, respectively, with $x$=0.2.}
\label{fig:dfbnt}
\end{figure}
\begin{figure}[t!]
\includegraphics[width=0.41\textwidth]{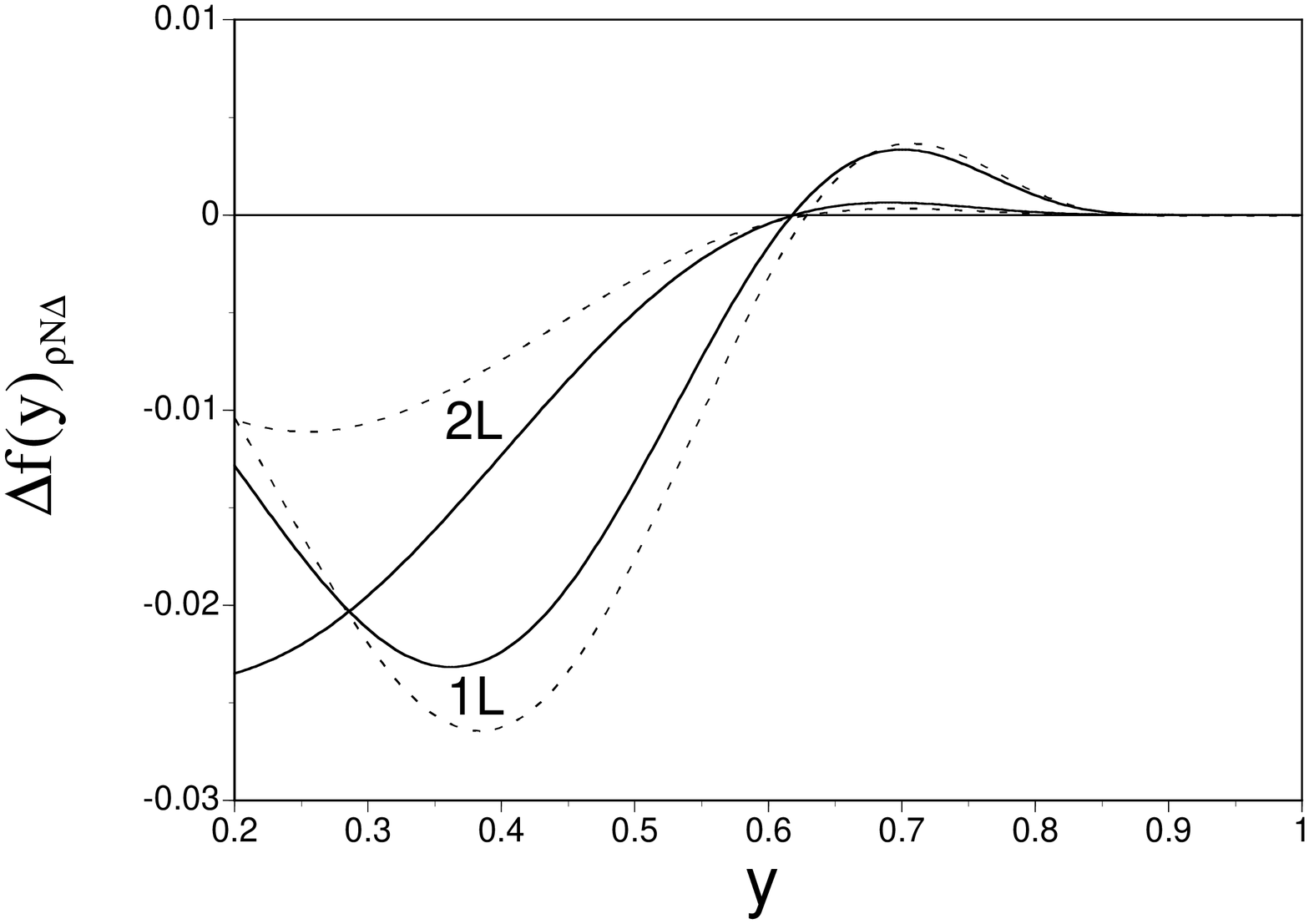}
\vspace{-0.8cm}
\caption{Meson momentum distributions $\Delta f_{1L}(y)$ 
         and $\Delta f_{2L}(y)$ from the $\rho N \Delta$ process.
         The solid and dashed curves are obtained at
         $Q^2$=1 and 2 GeV$^2$, respectively, with $x$=0.2.} 
\label{fig:dfbdl}
\vspace{0.7cm}
\includegraphics[width=0.41\textwidth]{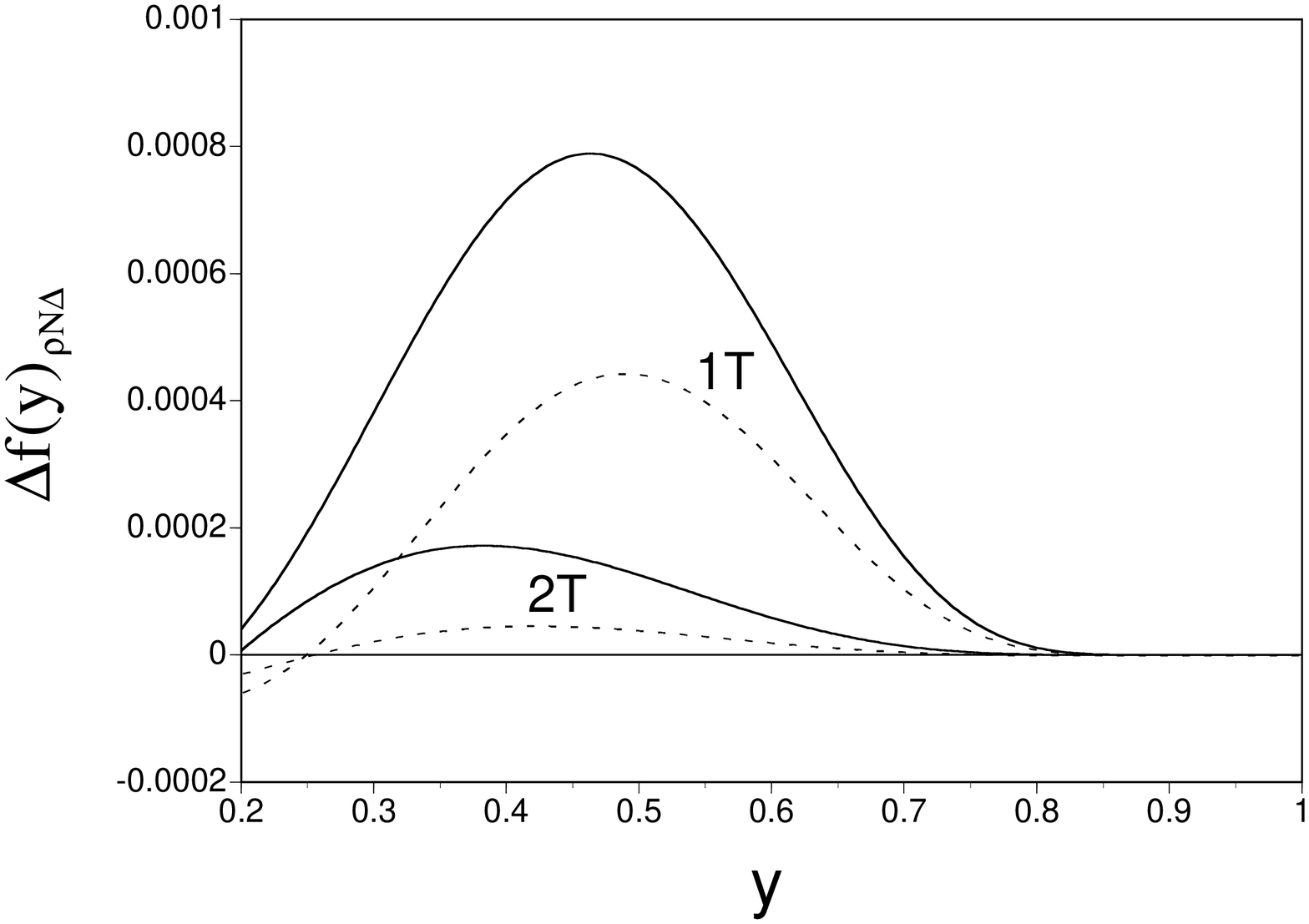}
\vspace{-0.8cm}
\caption{Meson momentum distributions $\Delta f_{1T}(y)$ 
         and $\Delta f_{2T}(y)$ from the $\rho N \Delta$ process.
         The solid and dashed curves are obtained at
         $Q^2$=1 and 2 GeV$^2$, respectively, with $x$=0.2.} 
\label{fig:dfbdt}
\end{figure}

We show the numerical results in Figs. \ref{fig:dfbnl}--\ref{fig:dfbdt}
for the vertex momentum (B), which is the preferred choice according
to Ref. \cite{julich}. In addition to the variable $y$,
the distributions depend also on $x$ and $Q^2$.
The distributions are calculated at $x=0.2$ with $Q^2$=1
and 2 GeV$^2$ for the solid and dashed curves, respectively.
Because of the $x$ dependence, the unphysical
region $y<x$ is not shown in these figures. 
In Figs. \ref{fig:dfbnl} and \ref{fig:dfbnt},
the distributions due to the $\rho NN$ process are shown.
In Figs. \ref{fig:dfbdl} and \ref{fig:dfbdt},
the distributions due to the $\rho N \Delta$ process are shown.
In these distributions, the isospin factors are taken out from
the distributions, so that the distributions
$\Delta f(y)/| \widetilde \phi_V^* \cdot \widetilde T |^2$
are actually shown, and there are differences of factors
3 (in $\rho NN$) and 
2 (in $\rho N \Delta$) from those in Ref. \cite{fs}.
The coupling constants are taken as
$g_V^2/(4\pi)=0.84$, $f_V=6.1 g_V$, and $f_{VN\Delta}^2/(4\pi)=20.45$
\cite{mhe}.

In spite of the positive $\Delta f_{1L}(y)$ and 
$\Delta f_{2L}(y)$ distributions from the $\rho NN$ process
in Fig. \ref{fig:dfbnl}, the distributions from the $\rho N\Delta$
are mainly negative. In the unpolarized case, the meson momentum
distributions are, of course, positive. However, because
the distribution $\Delta f(y)$ is defined by the helicity difference
in Eq. (\ref{eqn:dfy}), it becomes either positive or negative
depending on the helicity structure at the $VNB$ vertex.
For the meson angular momentum state $\ell_z$, the $VNB$ vertex
amplitude is proportional to $k_\perp^{\ell_z}$ and higher order
terms. Since the momentum $k_\perp$ is in general much smaller than
the nucleon mass, the vertex amplitudes with $\ell_z=0$
contribute dominantly to $\Delta f(y)$.
There is only one $\rho NN$ amplitude with $\ell_z=0$ and
$\lambda_V \ne 0$, and it has the helicity structure
$\lambda_V=+1$ and $\lambda_N'=-1/2$, where the initial  helicity
is fixed at $\lambda_N=+1/2$.
This fact indicates that $f_{1L}^{\lambda_V=+1}(y)$
and $f_{2L}^{\lambda_V=+1}(y)$ are certainly larger
than $f_{1L}^{\lambda_V=-1}(y)$ and $f_{1L}^{\lambda_V=-1}(y)$,
respectively, which results in the positive distributions
$\Delta f_{1L}(y)$ and $\Delta f_{1L}(y)$ from the $\rho NN$ process.
On the other hand, there are two amplitudes with $\ell_z=0$
and $\lambda_V \ne 0$ in the $\rho N \Delta$ process, and they have
helicity states $\lambda_V=-1$, $\lambda_\Delta=+3/2$ and 
$\lambda_V=+1$, $\lambda_\Delta=-1/2$.
Actually calculating these helicity amplitudes, we find that
both amplitudes depend much on the momentum choice,
namely (A) or (B), at the $\rho N \Delta$ vertex.
Therefore, $f_{1L}^{\lambda_V=+1}(y)$ and $f_{2L}^{\lambda_V=+1}(y)$
are either larger or smaller than $f_{1L}^{\lambda_V=-1}(y)$
and $f_{2L}^{\lambda_V=-1}(y)$, respectively, depending on
the momentum choice. In the prescription (B),
the positive helicity distributions $f_{1L}^{\lambda_V=+1}(y)$
and $f_{2L}^{\lambda_V=+1}(y)$ are mostly smaller, so that
$\Delta f_{1L}(y)$ and $\Delta f_{2L}(y)$ 
become negative distributions in the wide $x$ region.
However, the situation is opposite in the prescription (A),
where the distributions are mostly positive.

As expected, the distributions $\Delta f_{1L}(y)$
in Figs. \ref{fig:dfbnl} and \ref{fig:dfbdl} are the dominant
ones and they are almost independent of $Q^2$.
However, $\Delta f_{2L}(y)$, $\Delta f_{1T}(y)$,
and $\Delta f_{2T}(y)$ are roughly proportional to $1/Q^2$,
so that these new contributions become more important as $Q^2$
becomes smaller. Figures \ref{fig:dfbnl}$-$\ref{fig:dfbdt}
clearly show this tendency.
The transverse distributions $\Delta f_{1T}(y)$ and 
$\Delta f_{2T}(y)$ are an order of magnitude smaller than
the longitudinal ones $\Delta f_{1L}(y)$ and $\Delta f_{2L}(y)$.
Therefore, the major correction comes from the distribution
$\Delta f_{2L}(y)$, which is almost comparable magnitude with
$\Delta f_{1L}(y)$ in Figs. \ref{fig:dfbnl} and
\ref{fig:dfbdl}. Because the correction terms are proportional
to $\gamma^2 = 4 m_N^2 x^2 /Q^2$, they are small contributions
in the kinematical range $x<0.05$ with $Q^2>$1 GeV$^2$.
However, their effects become more pronounced as $x$ becomes
larger.

\section{Results}
\label{results}
\setcounter{equation}{0}

\begin{figure}[h!]
\includegraphics[width=0.41\textwidth]{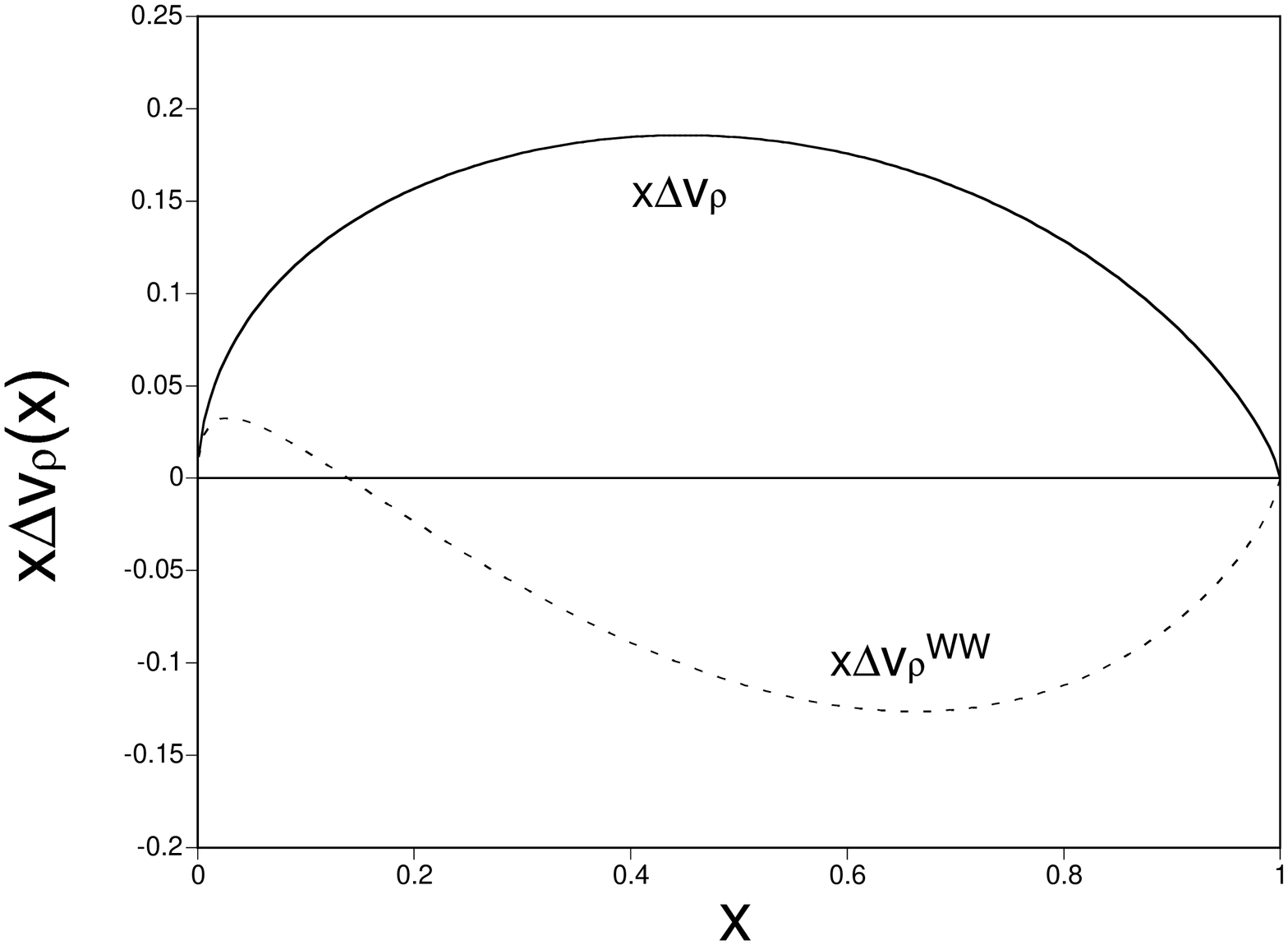}
\vspace{-0.4cm}
\caption{Assumed polarized valence-quark distribution in the $\rho$
         meson and the WW distribution
         at $Q^2$=1 GeV$^2$.}
\label{fig:dvrho}
\end{figure}

For calculating the $\Delta \bar u - \Delta \bar d$ distribution
numerically, we need the polarized antiquark distributions in $\rho$,
the isospin factors, and the vertex form factors.
The $\rho$-meson parton distributions are not known, so that
the same prescription is used as the one in Refs. \cite{fs,cs}.
Considering a lattice QCD estimate \cite{lattice}, the polarized
valence-quark distribution is assumed as
\begin{equation}
\Delta V_\rho (x,Q^2) = 0.6 \, V_\pi (x,Q^2) ,
\end{equation}
at $Q^2$=1 GeV$^2$. The distribution in the pion is taken from
the GRS (Gl\"uck, Reya, and Schienbein) parametrization
in 1999 \cite{grs99}. The charge symmetry suggests the relation
for the valence-quark distributions:
\begin{equation}
    (\Delta \bar u)_{\rho^-}^{val}
=   (\Delta \bar d)_{\rho^+}^{val}
= 2 (\Delta \bar u)_{\rho^0}^{val}
= 2 (\Delta \bar d)_{\rho^0}^{val}
= \Delta V_\rho .
\end{equation}
For the sea-quark distributions, they are assumed to be flavor symmetric.
Then, we obtain the $\Delta \bar u - \Delta \bar d$ distributions
in the $\rho$ meson:
\begin{align}
(\Delta \bar u - \Delta \bar d)_{\rho^+} & = - \Delta V_\rho ,
\nonumber \\
(\Delta \bar u - \Delta \bar d)_{\rho^0} & = 0 ,
\nonumber \\
(\Delta \bar u - \Delta \bar d)_{\rho^-} & = + \Delta V_\rho .
\label{eqn:rho-ud}
\end{align}
For the $g_2$ part of $\rho$, the Wandzura-Wilczek relation is used
as discussed in Sec. \ref{vector}:
\begin{equation}
\Delta V_\rho^{WW} (x,Q^2) 
     =  - \Delta V_\rho (x,Q^2)
       + \int_x^1 \frac{dy}{y} \Delta V_\rho (y,Q^2) .
\end{equation}
Both the valence-quark distribution and the WW distribution are
shown at $Q^2$=1 GeV$^2$ in Fig. \ref{fig:dvrho}.

\vspace{0.4cm}
\vspace{0.4cm}

Necessary isospin factors are calculated
from Eq. (\ref{eqn:isospin}) as
\begin{align}
| \! < \! n \, | 
      \, \widetilde \phi_{\rho^+}^* \! \cdot \widetilde T \, 
              | \, p \! > \! |^2  & = 2  \, ,
\nonumber \\
| \! < \! \Delta^0 \, | 
      \, \widetilde \phi_{\rho^+}^* \! \cdot \widetilde T \, 
              | \, p \! > \! |^2  & = 1/3 \, ,
\nonumber \\
| \! < \! \Delta^{++} \, | 
      \, \widetilde \phi_{\rho^-}^* \! \cdot \widetilde T \, 
              | \, p \! > \! |^2  & = 1  \, .
\label{eqn:isofactor}
\end{align}
Using Eqs. (\ref{eqn:dPi}), (\ref{eqn:rho-ud}), and (\ref{eqn:isofactor}),
we obtain
\begin{align}
(\Delta  &  \bar u   - \Delta \bar d )_{p \rightarrow \rho B}
 = \sum_{\rho, B} \bigg [ \, \big \{ \,
              \Delta f_{1L} +\Delta f_{1T} \, \big \}
          \otimes  (\Delta \bar u - \Delta \bar d)_{\rho}
\nonumber \\
& \ \ \ \ \ \ \ \ \ \ \ \ \ \ \ \ 
         - \big \{ \,
              \Delta f_{2L} +\Delta f_{2T} \, \big \}
          \otimes  (\Delta \bar u - \Delta \bar d)_{\rho}^{WW} \, \bigg ]
\nonumber \\
& 
= \bigg [ \, -2 \, \Delta f^{\rho NN}_{1L+1T} 
  + \frac{2}{3} \, \Delta f^{\rho N \Delta}_{1L+1T} \, \bigg ] 
          \otimes \Delta V_\rho
\nonumber \\
& \ 
- \bigg [ \, -2 \, \Delta f^{\rho NN}_{2L+2T} 
  + \frac{2}{3} \, \Delta f^{\rho N \Delta}_{2L+2T} \, \bigg ] 
          \otimes \Delta V_\rho^{WW}
\, ,
\label{eqn:ubdb}
\end{align}
where $\otimes$ indicates the convolution integral
in Eq. (\ref{eqn:dPi}):
\begin{equation}
a \otimes b = \frac{1}{1+\gamma^2}
             \int_x^1 \frac{dy}{y} \, a(y) \, b(x/y) .
\end{equation}
The meson momentum distributions $\Delta f^{\rho NN}_{1L+1T}$
and $\Delta f^{\rho NN}_{2L+2T}$ are defined by extracting 
the isospin factors:
\begin{equation}
\Delta f^{\rho NB}_{iL+iT} = \frac{ \Delta f_{iL} +\Delta f_{iT}}
   {| \widetilde \phi_{\rho}^* \! \cdot \widetilde T |^2},
\ \ \  \text{$i$=1, 2} .
\end{equation}
The expression of Eq. (\ref{eqn:ubdb}) may seem to be
different from Refs. \cite{fs,cs} even in the limit
$Q^2 \rightarrow \infty$; however, it is just the matter
of the definition of the meson momentum distributions.
They included the isospin factor 
\begin{equation}
| \! < \! n \, | 
      \, \widetilde \phi_{\rho^+}^* \! \cdot \widetilde T \, 
              | \, p \! > \! |^2 
 + | \! < \! p \, | 
      \, \widetilde \phi_{\rho^0}^* \! \cdot \widetilde T \, 
              | \, p \! > \! |^2 =3 \, ,
\end{equation}
in the distribution $\Delta f(y)$ for the $\rho NN$ process 
and the factor
\begin{align}
| \! < \! \Delta^{0} \, |   
      \, \widetilde \phi_{\rho^+}^* \! \cdot \widetilde T \, &
              | \, p \! > \! |^2
 + | \! < \! \Delta^{+} \, | 
      \, \widetilde \phi_{\rho^0}^* \! \cdot \widetilde T \, 
              | \, p \! > \! |^2
\nonumber \\
& + | \! < \! \Delta^{++} \, | 
      \, \widetilde \phi_{\rho^-}^* \! \cdot \widetilde T \, 
              | \, p \! > \! |^2  = 2
\end{align}
for the $\rho N \Delta$.
Therefore, our expression certainly agrees on those in
Refs. \cite{fs,cs} at $Q^2 \rightarrow \infty$.

The remaining quantities are the vertex form factors.
They are roughly known from the studies of one-boson-exchange
potentials (OBEPs); however, a slight change of the cutoff parameter
could result in a large difference of antiquark distributions.
Furthermore, there is an issue of the charge and momentum conservations
for the splitting process \cite{msm}
if a $t$ $(=(p_N-p_B)^2)$ dependent form factor
is used. A possible solution is to use the $t$ dependent form factor
multiplied by a $u$ dependent one \cite{zoller}.
For this purpose, it is more convenient to take an exponential
form factor so as to become the additional form $t+u$ within
the form factor:
\begin{equation}
F_{\rho NN}(k)=F_{\rho N \Delta}(k)
       = \exp \bigg [ \frac{m_N^2 -m_{VB}^2}{2\Lambda_e^2} \bigg ] ,
\end{equation}
where $m_{VB}^2$ is defined in Eq. (\ref{eqn:mvb2}), and 
the cutoff parameter $\Lambda_e$ is taken as $\Lambda_e$=1 GeV
in the following numerical results. In Ref. \cite{julich},
the cutoff parameters are obtained by fitting baryon-production
cross sections $pp \rightarrow BX$: $\Lambda_e^{\rho NN}$=1.10 GeV
and $\Lambda_e^{\rho N \Delta}$=0.98 GeV. However, the parameters
are not well determined in general.
We discuss the dependence on this cutoff value at the end 
of this section. The form factors are the same as the ones
in the previous publications \cite{fs,cs},
so that we could compare our results with theirs.

\begin{figure}[t!]
\includegraphics[width=0.41\textwidth]{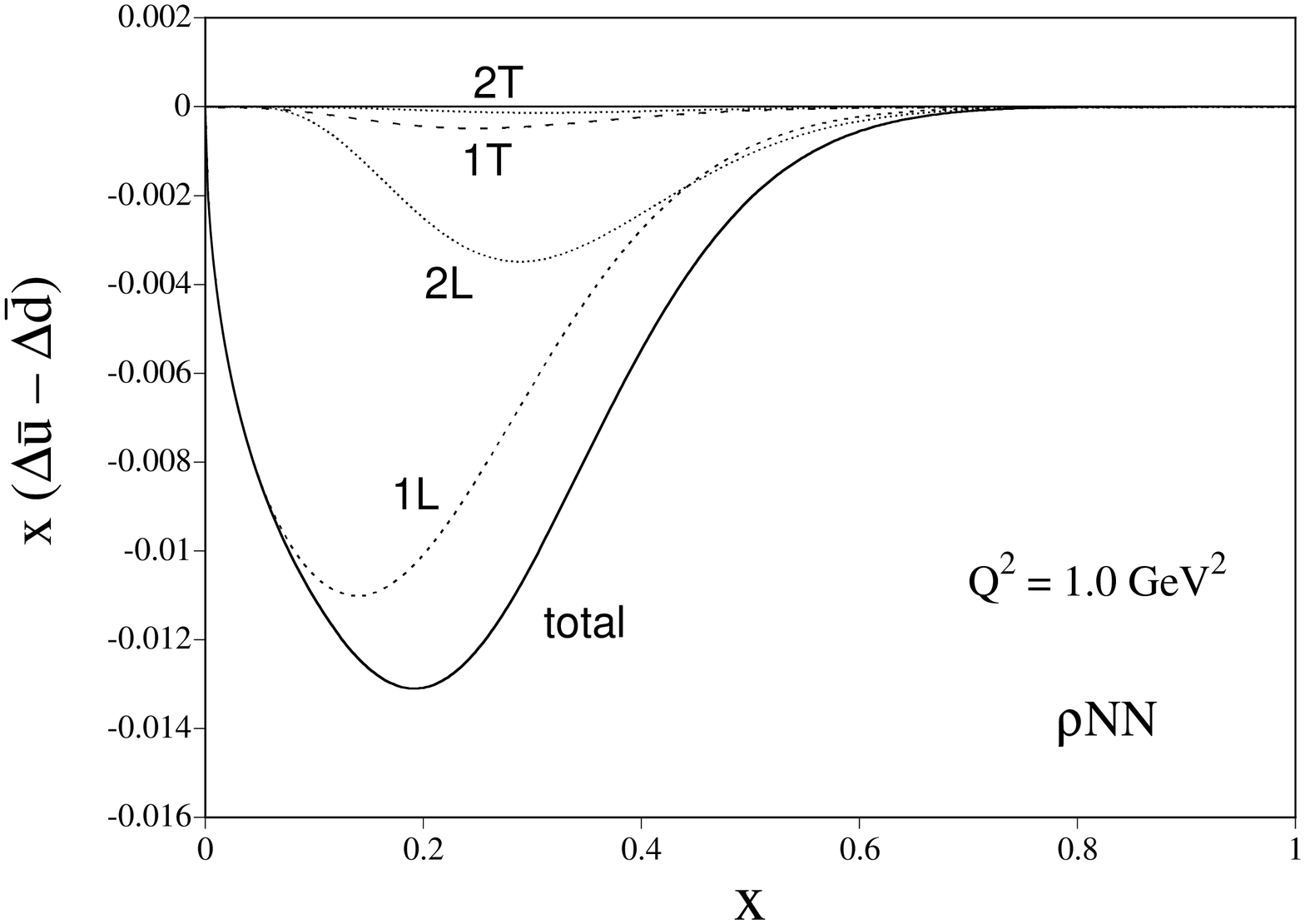}
\vspace{-0.4cm}
\caption{$\Delta \bar u - \Delta \bar d$ distributions
         from the $\rho NN$ process at $Q^2$=1 GeV$^2$.
         The $1L$, $2L$, $1T$, and $2T$ type contributions
         and their summation are shown.}
\label{fig:ubdbn}
\vspace{0.9cm}
\includegraphics[width=0.41\textwidth]{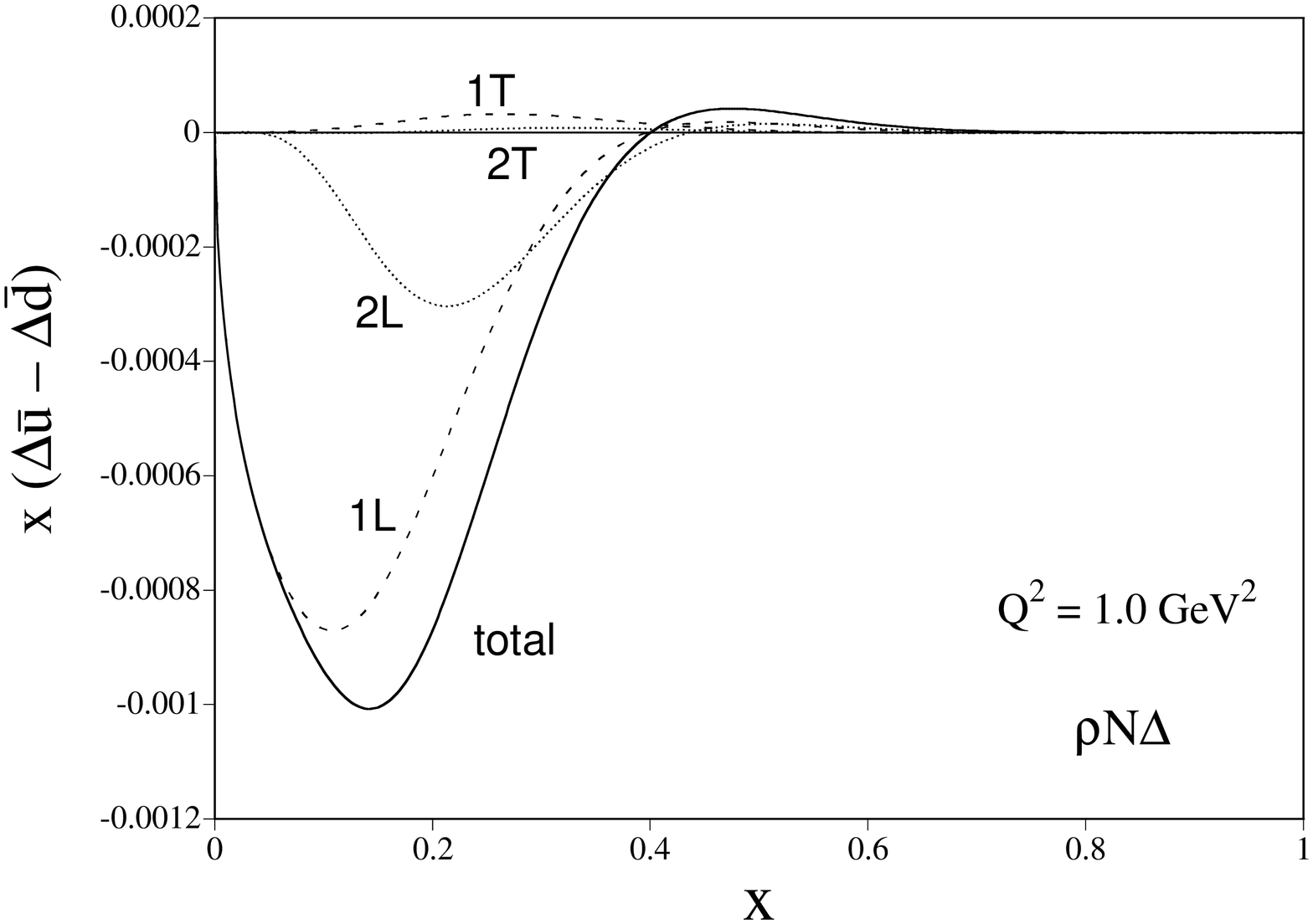}
\vspace{-0.4cm}
\caption{$\Delta \bar u - \Delta \bar d$ distributions
         from the $\rho N \Delta$ process at $Q^2$=1 GeV$^2$.
         The $1L$, $2L$, $1T$, and $2T$ type contributions
         and their summation are shown.}
\label{fig:ubdbd}
\vspace{0.9cm}
\includegraphics[width=0.41\textwidth]{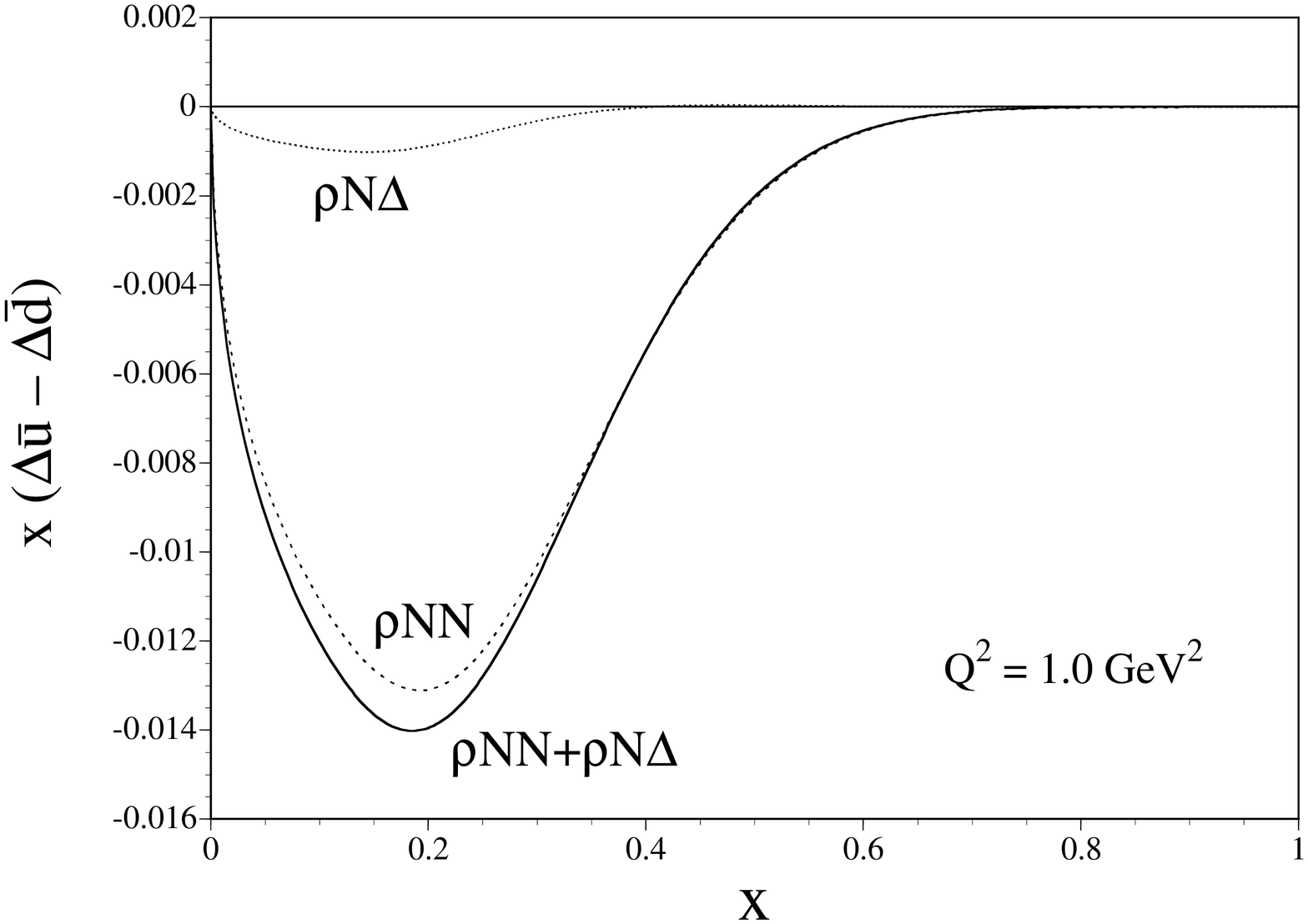}
\vspace{-0.4cm}
\caption{$\Delta \bar u - \Delta \bar d$ distributions
         from the $\rho NN$ and $\rho N \Delta$ processes
         at $Q^2$=1 GeV$^2$.}
\label{fig:ubdbnd}
\end{figure}

Using these form factors and the parton distributions in $\rho$,
we obtain the $\rho NN$ and $\rho N \Delta$ process contributions
to the $\Delta \bar u - \Delta \bar d$ in the nucleon.
In Fig. \ref{fig:ubdbn}, the $1L$, $2L$, $1T$, and $2T$ type
distributions from the $\rho NN$ process are shown 
at $Q^2$=1 GeV$^2$ together with their total. 
The ordinary $1L$ term is the dominant contribution; however,
the $2L$ term becomes important at $x>0.3$. It is as large as
the $1L$ distribution in the medium $x$ region although
it is fairly small at $x<0.05$. The $1T$ and $2T$ distributions
are very small in the whole $x$ range.
Because $p \rightarrow \rho^+ n$ is the only contributing process
in which the valence $\bar d$ distribution in $\rho^+$ plays
the main role, the $\rho NN$ contributions are negative
in the $\Delta \bar u - \Delta \bar d$ in the nucleon.

Each term contribution has almost the same tendency in
the $\rho N \Delta$ process as shown in Fig. \ref{fig:ubdbd}:
the $1L$ term is the major one and the $2L$ term provides
some corrections depending on the $x$ region. 
There are two contributing processes, $p \rightarrow \rho^+ \Delta^0$
and $p \rightarrow \rho^- \Delta^-$, and the isospin factor is
three times larger in the latter one. This fact may seem to indicate
that the $\rho N \Delta$ processes provide a positive contribution
to $\Delta \bar u - \Delta \bar d$ in the nucleon due to
the valence $\bar u$ distribution in $\rho^-$. This kind of
explanation is certainly valid in the unpolarized flavor asymmetry
\cite{udsum,skubdb}. However, this is not the case in Fig. \ref{fig:ubdbd},
where the $1L$ and $1T$ distributions are mostly negative.
This misleading result comes from the helicity structure at
the $N \rightarrow \rho \Delta$ vertex. Although the helicity
difference $\Delta f(y)$ is positive for the $\rho NN$, it
is negative for the $\rho N \Delta$ in the case (B)
as explained in Sec. \ref{meson}.
Therefore, the $\rho N \Delta$ contribution becomes also
negative for the $\Delta \bar u - \Delta \bar d$ distribution. 

Next, the $\rho NN$ and $\rho N \Delta$ contributions are compared
in Fig. \ref{fig:ubdbnd}. The magnitude of the $\rho N \Delta$
contribution is very small compared with the $\rho N N$ one
in (B). From Figs. \ref{fig:dfbnl} and \ref{fig:dfbdl}, we find
that the magnitude of $\Delta f^{\rho N\Delta}(y)$ is already
three times smaller than $\Delta f^{\rho NN}(y)$,
and the $\rho N\Delta$ contribution
is further suppressed by the isospin factor (2/3)/2=1/3.
Therefore, the overall magnitude becomes much smaller.

\begin{figure}[t!]
\includegraphics[width=0.41\textwidth]{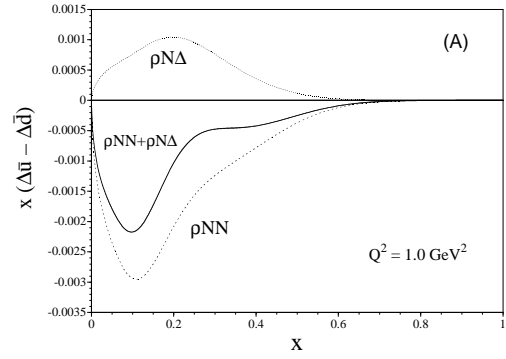}
\vspace{-0.6cm}
\caption{$\Delta \bar u - \Delta \bar d$ distributions
         from the $\rho NN$ and $\rho N \Delta$ processes
         at $Q^2$=1 GeV$^2$ for the vertex momentum choice (A).}
\label{fig:ubdbnd-a}
\end{figure}

As discussed in Sec. \ref{meson}, we may have another vertex choice
(A) instead of (B). In showing the numerical results so far,
the model (B) has been used. We show the choice (A) results in Fig.
\ref{fig:ubdbnd-a}. It is obvious that the distributions depend
much on this vertex choice. There are two major differences from Fig.
\ref{fig:ubdbnd}. One is that the order of magnitude is much smaller
in the $\rho NN$ distribution, and the other is that
the $\rho N \Delta$ distribution becomes positive. 
These are due to the difference of helicity structure at the vertices
between (A) and (B).

We also discuss the vertex cutoff dependence. 
The vertex cutoff has been taken as $\Lambda_e$=1 GeV in this section;
however, it is well known that calculated antiquark distributions are
very sensitive to the cutoff value \cite{skubdb}.
In the present paper, the exponential form factor is used
instead of dipole or monopole form factor, which is more 
popular in the studies of OBEPs.
The cutoff parameters of different form factors
could be related by \cite{skubdb}
\begin{equation}
\Lambda_1 =0.62 \Lambda_2 = 0.78 \sqrt{2} \Lambda_e ,
\label{eqn:cut12e}
\end{equation}
where the monopole and dipole parameters are defined
by the form factors
\begin{equation}
F^{(1)}_{VNB}(k)  = \frac{1-m_N^2/\Lambda_1^2}{1-m_{VB}^2/\Lambda_1^2} ,
\ \ 
F^{(2)}_{VNB}(k)  = \frac{(1-m_N^2/\Lambda_2^2)^2}
                          {(1-m_{VB}^2/\Lambda_2^2)^2} .
\end{equation}

\begin{figure}[t!]
\includegraphics[width=0.41\textwidth]{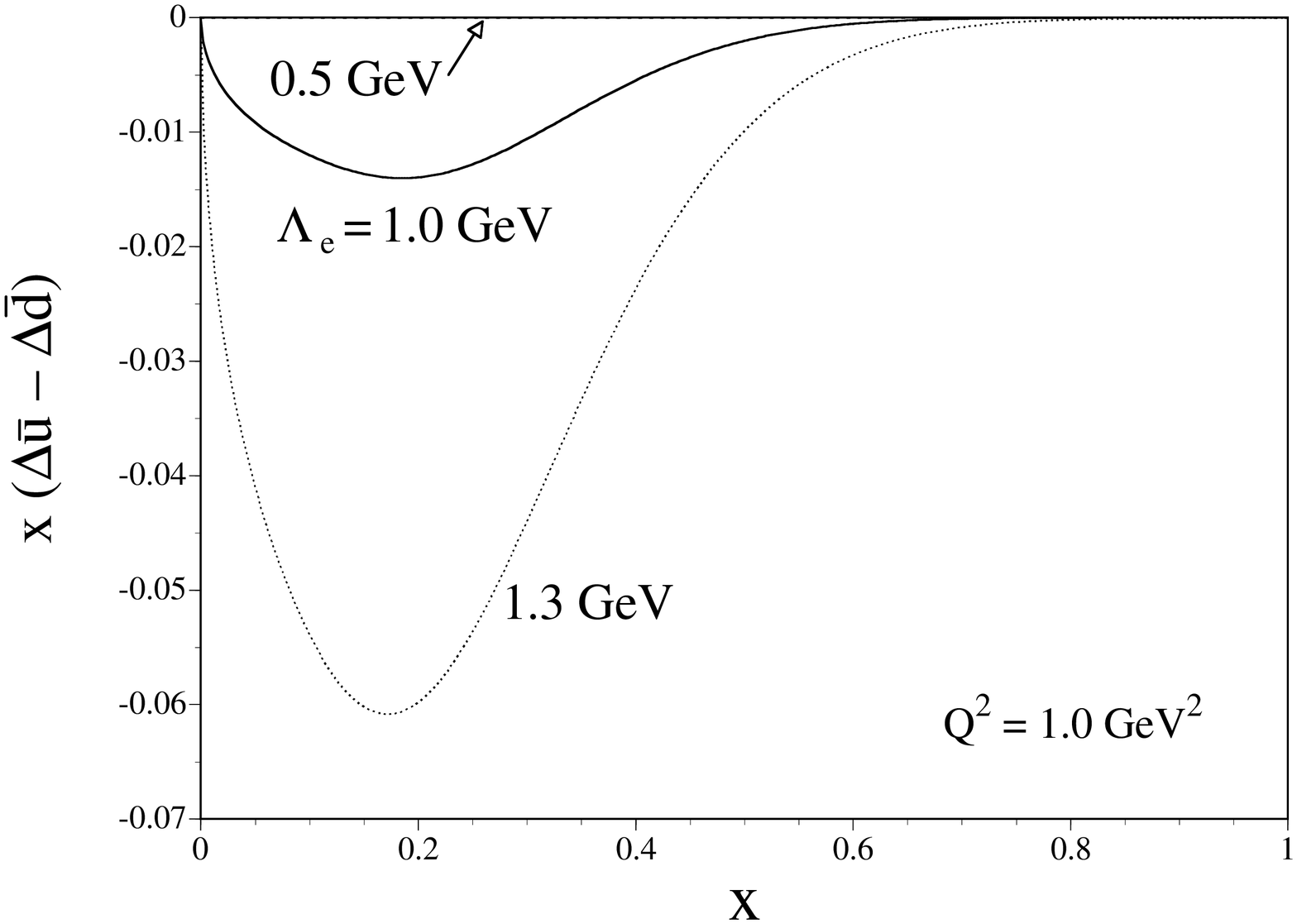}
\vspace{-0.5cm}
\caption{Cutoff dependence of $\Delta \bar u - \Delta \bar d$
         is shown at $Q^2$=1 GeV$^2$
         by taking $\Lambda_e$=0.5, 1.0, and 1.3 GeV.}
\label{fig:cutx}
\vspace{0.6cm}
\includegraphics[width=0.41\textwidth]{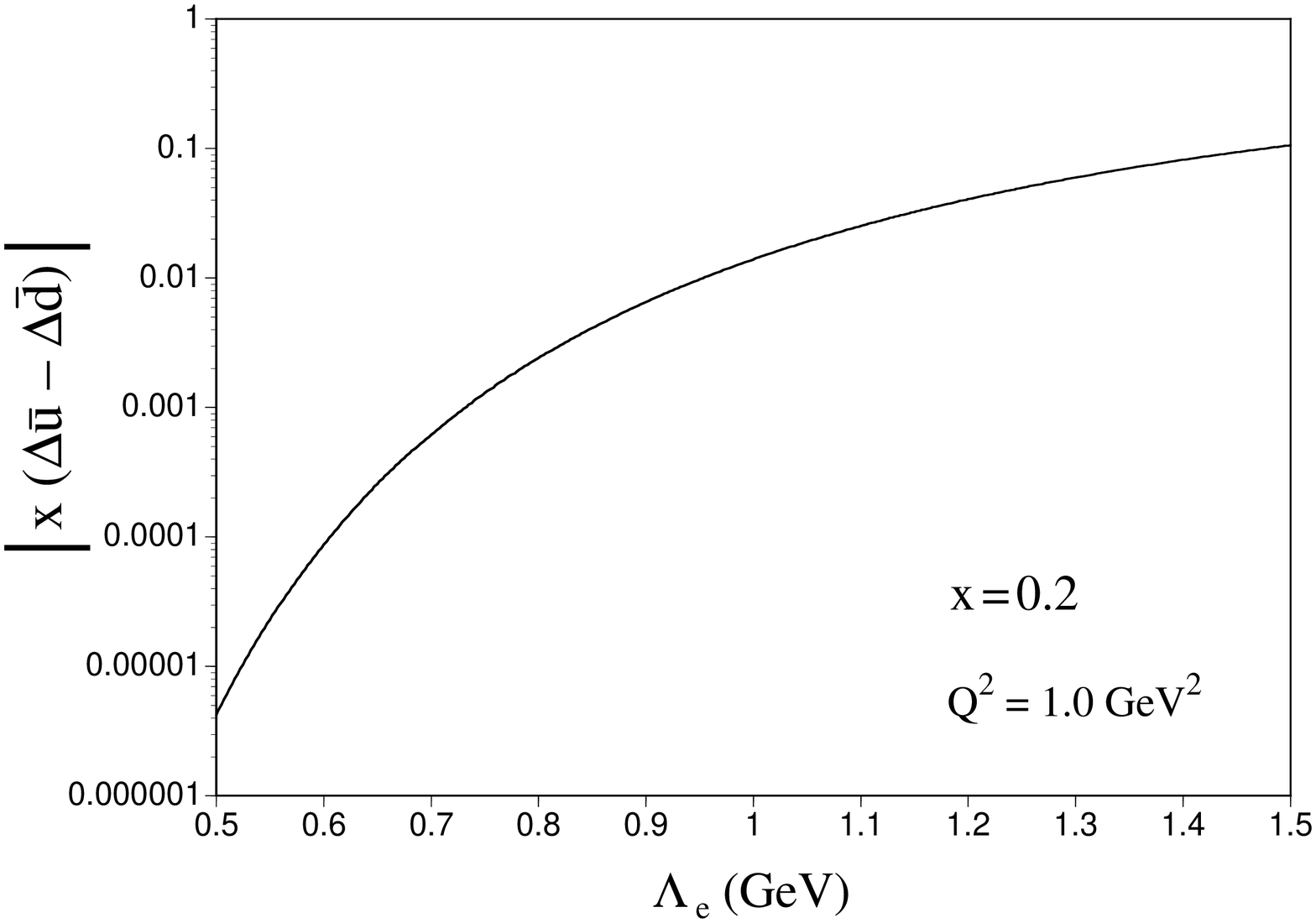}
\vspace{-0.5cm}
\caption{Cutoff dependence of $| \Delta \bar u - \Delta \bar d |$
         is shown at $x=0.2$ and $Q^2$=1 GeV$^2$
         as a function of the parameter $\Lambda_e$.}
\label{fig:cutdep}
\end{figure}

Even in the well investigated pion-nucleon coupling, the monopole
cutoff parameter $\Lambda_1$ ranges from 0.6 GeV to 1.4 GeV in quark
models and OBEPs \cite{skubdb}. It roughly corresponds to
$\Lambda_e$ from 0.5 GeV to 1.3 GeV according to
Eq. (\ref{eqn:cut12e}). We show the $\Delta \bar u - \Delta \bar d$
distributions for various cutoff parameters,
$\Lambda_e$=0.5, 1.0, and 1.3 GeV in Fig. \ref{fig:cutx}
for the prescription (B).
In the unpolarized distribution $\bar u - \bar d$ in Ref. \cite{skubdb},
it is fortunate that the cutoff dependence is rather small
because of the cancellation between the $\pi NN$ and $\pi N \Delta$
processes. However, the $\rho NN$ is the dominant contribution,
as shown in Fig. \ref{fig:ubdbnd}, in the present polarized studies
of (B), so that the overall magnitude is much dependent on the cutoff
parameter. There are orders of magnitude differences between
the three curves. 
Next, fixing $x$ at 0.2, we show the cutoff dependence in
Fig. \ref{fig:cutdep}. In fact, there is four orders of magnitude
variation from $\Lambda_e$=0.5 to 1.5 GeV. 
Therefore, an accurate determination of the cutoff parameters
is a key for a reliable meson-cloud prediction.

There have been studies in chiral soliton models \cite{models}.
They predict very different distributions, namely
$\Delta \bar u$ excess over $\Delta \bar d$ and
the order of magnitude of $\Delta \bar u - \Delta \bar d$
is large ($\sim 0.4$) at $x$=0.2. Although the soliton models
and the meson-cloud models obtain very similar distributions
for the unpolarized $\bar u-\bar d$, it is interesting to find
opposite prediction for the polarized distributions. 
The physics reason for the difference is not clear at this stage.
In any case, the distribution $\Delta \bar u - \Delta \bar d$ 
should be clarified experimentally in the near future by
$W$ production process \cite{rhic} and semi-inclusive experiments
\cite{compass}. Furthermore, there is a possibility to use the
polarized proton-deuteron Drell-Yan process \cite{pddy}
in combination with the proton-proton reaction.
Until the data will be taken, the theoretical predictions should
be discussed in details for comparison.

\section{Conclusions}
\label{concl}

The $\rho$ meson contributions to the polarized antiquark distribution
$\Delta \bar u - \Delta \bar d$ have been investigated.
In particular, we pointed out that the $g_2$ part of $\rho$ contributes
to the polarized distributions in the nucleon. We obtained the extra
contributions denoted as $2L$, $1T$, and $2T$ in addition to
the ordinary one ($1L$). Although the extra terms
are small in the small $x$ region ($x<0.05$), the magnitude of
the $2L$ term becomes comparable to the ordinary one
in the $x$ region $x>0.2$. 
The $g_2^\rho$ contributions are important in the kinematical region
of medium $x$ with small $Q^2$. The obtained $\Delta \bar u - \Delta \bar d$
is very sensitive to the cutoff parameter. The model should be investigated
further in order to compare with future experimental data.

\begin{acknowledgments}
S.K. and M.M. were supported by the Grant-in-Aid for Scientific Research
from the Japanese Ministry of Education, Culture, Sports, Science,
and Technology. M.M. was also supported by the JSPS Research Fellowships
for Young Scientists.
They would like to thank R. J. Fries for communications about
the calculations in Refs. \cite{fs,cs,julich,mt}.
\end{acknowledgments}

\appendix
\section{Analytical expressions of meson momentum distributions}
\label{appen}
\setcounter{equation}{0}

In the limit $Q^2 \rightarrow \infty$, the following meson momentum
distributions agree on those by Fries and Sch\"afer (FS) \cite{fs}
with a minor misprint in a $\rho N \Delta$ term. 
The situation of the momentum distributions is somewhat confusing
in the sense that Cao and Signal (CS) \cite{cs} pointed out two major
differences from Ref. \cite{fs} although the formalism is exactly
the same except for interference terms. According to Ref. \cite{cs}, 
all the $g_{\rho NN} f_{\rho NN}$ ($g_v f_v$ in our notation)
terms should be replaced by $-g_{\rho NN} f_{\rho NN}$,
and the momentum (B) results for $\rho N \Delta$ agree on
those of (A) in Ref. \cite{fs} instead of (B).

In spite of their claim, we believe that the FS results are right
with the following reasons. We also checked the helicity amplitudes
in Ref. \cite{julich}, which is referred to as J\"ulich in the following.
In addition to obvious typos, our results differ from the J\"ulich
expressions. First, complex conjugate should be taken if their expressions
are given for the process $N \rightarrow \rho B$ as indicated in
their appendix. Second, $f_v$ terms have different sign.
If the J\"ulich amplitudes were written for $N \rho \rightarrow B$ or
$B \rightarrow \rho N$, we would agree on their expressions.
Depending on the $\rho$ momentum direction, the $f_v$ term in
Eq. (\ref{eqn:vvnn}) becomes either positive or negative,
which could lead to the different sign of the $g_v f_v$ terms.
However, it is obvious that the outgoing meson is considered
in the formalism. Furthermore, taking summations of the helicity
amplitudes, we reproduce the unpolarized momentum distributions of 
Melnitchouk and Thomas \cite{mt,fries}, whereas the CS results
are inconsistent. The $VNN$ vertex in Eq. (\ref{eqn:vvnn})
is also consistent with the one in Ref. \cite{mach86}.

We also tested $\rho N \Delta$ helicity amplitudes in the vertex
momentum (B), but the results disagree on the J\"ulich expressions.
However, if the momentum (A) is used, our results agree on them.
It seems to us that the helicity amplitudes are shown
for the choice (B) in $\rho NN$ and for (A) in $\rho N \Delta$.
Therefore, as far as we investigated, we believe that
the FS calculations are right also for the $\rho N \Delta$ process.

In the following, we show the helicity dependent meson momentum
distributions. Because the distributions with $\lambda_V=0$
are irrelevant for calculating $\Delta f(y)$, so that they are not shown.
The isospin factors are extracted out from the expressions.

\begin{align}
& f^{\lambda_V}_{1L,\, VNB}\, (y) 
 =  \int_0^{(\vec k_\perp^{\, 2})_{max}} \frac{d\kp2}{16\pi^2}
  \frac{F^2_{VNB}\, (\yd,\, \kp2)}
      {\left\{ \mn^2 - m_{VB}^2\, (\yd,\, \kp2) \right\}^2}
  \nonumber \\ & \ \ \times
  \frac{\partial \yd}{\partial y}\, \frac{y}{\yd}
  \left\{ 1 + \frac{\kp2}{y\yd \mn^2} 
  ( \sqrt{1+\gamma^2} - 1) \right\} \, 
  D^{L,\, \lambda_V}_{VNB}\, (\yd,\, \kp2) \ ,
\end{align}
\vspace{-0.2cm}
\begin{align}
& f^{\lambda_V}_{1T,\, VNB}\, (y) 
 = \int_0^{(\vec k_\perp^{\, 2})_{max}} \frac{d\kp2}{16\pi^2}
  \frac{F^2_{VNB}\, (\yd,\, \kp2)}
       {\left\{ \mn^2 - m_{VB}^2\, (\yd,\, \kp2) \right\}^2}
  \nonumber \\ & \ \ \times
  \frac{\partial \yd}{\partial y}\, \frac{y}{\yd} \,
  \gamma^2 \, \frac{\kperp}{2 y \mn} \,
  D^{T,\, \lambda_V}_{VNB}\, (\yd,\, \kp2) \ ,
\end{align}
\vspace{-0.2cm}
\begin{align}
& f^{\lambda_V}_{2L,\, VNB}\, (y) 
 = \int_0^{(\vec k_\perp^{\, 2})_{max}} \frac{d\kp2}{16\pi^2}
  \frac{F^2_{VNB}\, (\yd,\, \kp2)}
       {\left\{ \mn^2 - m_{VB}^2\, (\yd,\, \kp2) \right\}^2} 
  \nonumber \\ & \ \ \times
  \frac{\partial \yd}{\partial y}\, \frac{y}{\yd} \,
  \gamma^2 \, \frac{\mv^2}{y^2 \mn^2} \,
  D^{L,\, \lambda_V}_{VNB}\, (\yd,\, \kp2) \ ,
\end{align}
\vspace{-0.2cm}
\begin{align}
& f^{\lambda_V}_{2T,\, VNB}\, (y) 
 = \int_0^{(\vec k_\perp^{\, 2})_{max}} \frac{d\kp2}{16\pi^2}
  \frac{F^2_{VNB}\, (\yd,\, \kp2)}
       {\left\{ \mn^2 - m_{VB}^2\, (\yd,\, \kp2) \right\}^2} 
  \nonumber \\ & \ \ \times
  \frac{\partial \yd}{\partial y}\, \frac{y}{\yd} \,
  \gamma^2 \, \frac{\kperp  \mv^2}{2 y^2 \yd m_N^3}
  ( \sqrt{1+\gamma^2} -1 ) \, 
  D^{T,\, \lambda_V}_{VNB}\, (\yd,\, \kp2) \ .
\end{align}
Here, the partial derivative is given by
\begin{equation}
\frac{\partial \yd}{\partial y}\, \frac{y}{\yd} =
\left\{ 1 + \frac{\gamma^2}{y^2 \mn^2} (\kp2+\mv^2) \right\}^{-1/2} .
\end{equation}

The distributions $D^{L,\, \lambda_V}_{VNN} \, (\yd,\, \kp2)$
are calculated for the prescription (A) as
\begin{align}
& D^{L,\, +1}_{VNN}\, (\yd,\, \kp2)
 = \frac{2}{\yp{3} (1-\yd)^2} \,
  \bigg[ \, g_V^2 \left( \kp2 + \yp{4} \mn^2 \right)
  \nonumber \\ & \ \
  + g_V f_V \yd \left[ \kp2
    + \yd \left\{ \yp{2}\mn^2 - (1-\yd) \mv^2 \right\} \right]
  \nonumber \\ & \ \
  + \frac{f_V^2}{4\mn^2} \left[ \yp{2}\mn^2 \kp2
  + \left\{ \yp{2}\mn^2 - (1-\yd)\mv^2 \right\}^2 \right] \bigg] \, ,
\end{align}
\begin{align}
& D^{L,\, -1}_{VNN}\, (\yd,\, \kp2) 
 = \frac{2 \kp2}{\yp{3}(1-\yd)^2} \,
  \bigg[ \, g_V^2 (1-\yd)^2 
  \nonumber \\ & \ \
  - g_V f_V \yd (1-\yd)
   + \frac{f_V^2}{4\mn^2}
  \left( \kp2 + \yp{2}\mn^2 \right) \bigg] \, ,
\end{align}
\begin{align}
& D^{T,\, \lambda_V}_{VNN}\, (\yd,\, \kp2)
 = \frac{\lambda_V \kperp}{\yp{3}(1-\yd)^2} \,
  \bigg[ \, -2 g_V^2 \yp{2}(1-\yd)\mn
  \nonumber \\ & \ \
  + \frac{g_V f_V}{\mn} \left\{ \kp2 - \yp{2}(1-2\yd)\mn^2
  + (1-\yd)^2\mv^2 \right\}
  \nonumber \\ & \ \
  + \frac{f_V^2}{4\mn^2} 2\yd\mn
  \left\{ \kp2 + \yp{2}\mn^2 - (1-\yd)\mv^2 \right\} \bigg] \, .
\end{align}

In the $VN\Delta$ process, the distributions are calculated for the
vertex momentum (A) as
\begin{align}
& D^{L,\, +1}_{VN\Delta}\, (\yd,\, \kp2) 
 = \frac{f_{VN\Delta}^2}{3 \yp{3} (1-\yd)^4 \md^2 \mv^2} \,
  \nonumber \\ & \ \ 
\times   \bigg[ \, \kperp^6 
+ \kperp^4 \left\{ 3 - 4\yd(1-\yd) \right\}\md^2
  \nonumber \\ & \ \ 
+ \kp2 \Big[ \yp{2} \left\{ 2\yp{2} + (2-\yd)^2 \right\} \md^4
  \nonumber \\ & \ \
  + 4\yd(1-\yd)^3\mn\md\mv^2 \Big]
  + \yp{4}\md^6
  \nonumber \\ & \ \
  - 2\yp{2}(1-\yd)^3\mn\md^3\mv^2 + (1-\yd)^6\mn^2\mv^4 \bigg] \, ,
\end{align}
\begin{align}
& D^{L,\, -1}_{VN\Delta}\, (\yd,\, \kp2) 
 = \frac{f_{VN\Delta}^2}{3 \yp{3} (1-\yd)^2 \md^2 \mv^2} \,
\nonumber \\ & \ \
\times \bigg[ \, \kperp^4 \mn^2 + \kperp^2 \{ 4\yp{2}\mn^2\md^2 
    - 4\yd(1-\yd)\mn\md\mv^2
\nonumber \\ & \ \
   +(1-\yd)^2\mv^4  \} + 3\yp{4}\mn^2\md^4 
- 6 \yp{2}(1-\yd)\mn\md^3\mv^2
\nonumber \\ & \ \
 + 3(1-\yd)^2\md^2\mv^4 \bigg] \, ,
\end{align}
\begin{align}
& D^{T,\, \lambda_V}_{VN\Delta}\, (\yd,\, \kp2) 
 = \frac{\lambda_V f_{VN\Delta}^2 \kperp}
  {3 \yp{3} (1-\yd)^3 \md^2 \mv^2} \,
  \bigg[ \, \kperp^4 \mn
  \nonumber \\ & \ \
  - 2\yd\kperp^2 \left\{ (1-2\yd)\mn\md^2
  + (1-\yd)\md\mv^2 \right\}
  \nonumber \\ & \ \
  - (2-3\yd)\yp{3}\mn\md^4 - 2\yp{2}(1-\yd)\md^3\mv^2
  \nonumber \\ & \ \
  + 2\yd(1-\yd)^3\mn^2\md\mv^2 - (1-\yd)^4\mn\mv^4 \bigg] \, .
\end{align}

In the same way, the distributions are obtained for the prescription
(B) as
\begin{align}
& D^{L,\, +1}_{VNN}\, (\yd,\, \kp2) 
 = \frac{2}{\yp{3} (1-\yd)^2} \,
  \bigg[ \, g_V^2 \left( \kp2 + \yp{4} \mn^2 \right)
  \nonumber \\ & \ \
   + g_V f_V \yd \left\{ (1+\yd)\kp2 + 2\yp{3}\mn^2 \right\}
  \nonumber \\ & \ \
   + \frac{f_V^2}{4\mn^2} \left( \kp2 + \yp{2}\mn^2 \right)
  \left( \kp2 + 4\yp{2}\mn^2 \right) \bigg] \, ,
\end{align}
\begin{align}
& D^{L,\, -1}_{VNN}\, (\yd,\, \kp2)
 = \frac{2 \kp2}{\yp{3} (1-\yd)^2} \,
  \bigg[ \, g_V^2 (1-\yd)^2 
  \nonumber \\ & \ \
    - g_V f_V \yd (1-\yd) + \frac{f_V^2}{4\mn^2}
  \left( \kp2 + \yp{2}\mn^2 \right) \bigg] \, ,
\end{align}
\begin{align}
& D^{T,\, \lambda_V}_{VNN}\, (\yd,\, \kp2) 
 = \frac{\lambda_V \kperp}{\yp{3} (1-\yd)^2} \,
  \bigg[ \, -2 g_V^2 \yp{2}(1-\yd)\mn
  \nonumber \\ & \ \
  + \frac{g_V f_V}{\mn} \yd \left\{ \kp2 - (2-3\yd)\yd\mn^2 \right\}
  \nonumber \\ & \ \
  + \frac{f_V^2}{4\mn^2} 4\yd\mn
  \left\{ \kp2 + \yp{2}\mn^2 \right\} \bigg] \, ,
\end{align}

\begin{align}
& D^{L,\, +1}_{VN\Delta}\, (\yd,\, \kp2) 
 = \frac{f_{VN\Delta}^2}{3 \yp{3} (1-\yd)^4 \md^2 \mv^2} \,
  \bigg[ \, \kperp^6 
  \nonumber \\ & \ \
  + \kperp^4 \left\{ (3-4\yd+4\yp{2})\md^2 \right.
  - 4 \yd(1-\yd)^2\mn\md 
  \nonumber \\ & \ \
\left. + (1-\yd)^4\mn^2 \right\}
  + \kp2 \left\{ \yp{2}(4-4\yd+3\yp{2})\md^4 \right.
  \nonumber \\ & \ \
  - 2\yp{2}(1-\yd)^2\mn\md^3 
  + 2\yd(1-\yd)^4\mn^2\md^2 
  \nonumber \\ & \ \
+ 4\yp{2}(1-\yd)^3\mn^3\md
   - \left. 2 \yd(1-\yd)^5\mn^4 \right\} + \yp{4}\md^6
  \nonumber \\ & \ \
  + 2\yp{3}(1-\yd)^2\mn\md^5 
  + \yp{2}(1-\yd)^4\mn^2\md^4  
  \nonumber \\ & \ \
- 2\yp{3}(1-\yd)^3\mn^3\md^3
  - 2\yp{2}(1-\yd)^5\mn^4\md^2 
  \nonumber \\ & \ \
+ \yp{2}(1-\yd)^6\mn^6 \bigg] \, ,
\end{align}
\begin{align}
& D^{L,\, -1}_{VN\Delta}\, (\yd,\, \kp2)
 = \frac{f_{VN\Delta}^2}{3 \yp{3} (1-\yd)^2 \md^2 \mv^2} \,
  \bigg[ \, \kperp^6 
  \nonumber \\ & \ \
+ \kperp^4 \left\{ (3+2\yd)\md^2 \right.
   + \left. 4\yd\mn\md +(1-2\yd+2\yp{2})\mn^2 \right\}
  \nonumber \\ & \ \ 
   + \kp2 \left\{ \yd(6+\yd)\md^4 +10\yp{2}\mn\md^3 \right.
  \nonumber \\ & \ \
   - 2\yd(3-4\yd-\yp{2})\mn^2\md^2 - 4\yp{2}(1-\yd)\mn^3\md
  \nonumber \\ & \ \
   + \left. \yp{2}(1-\yd)^2\mn^4 \right\} + 3\yp{2}\md^6
   + 6\yp{3}\mn\md^5
  \nonumber \\ & \ \
   - 3 \yp{2}(2-2\yd-\yp{2})\mn^2\md^4 - 6\yp{3}(1-\yd)\mn^3\md^3
  \nonumber \\ & \ \
   + 3\yp{2}(1-\yd)^2\mn^4\md^2 \bigg] \, ,
\label{eqn:dld-1}
\end{align}
\begin{align}
& D^{T,\, \lambda_V}_{VN\Delta}\, (\yd,\, \kp2) 
 = \frac{\lambda_V f_{VN\Delta}^2 \kperp}
  {3 \yp{2} (1-\yd)^3 \md^2 \mv^2} 
  \nonumber \\ & \ \
  \times \bigg[ \, \kperp^4 \left\{ 2\md + (2-\yd)\mn \right\}
  + \kp2 \left\{ 4\yd\md^3 \right.
  \nonumber \\ & \ \
- 2(2-4\yd+\yp{2})\mn\md^2 
  - 2(1-\yd)\mn^2\md 
  \nonumber \\ & \ \
\left. + 2(1-\yd)^3\mn^3 \right\} 
  + 2\yp{2}\md^5 - \yd(1-2\yp{2})\mn\md^4 
  \nonumber \\ & \ \
 - 2\yd(1-\yd)\mn^2\md^3 + 2\yd(1-\yd)^3\mn^3\md^2 
  \nonumber \\ & \ \
+ 2\yd(1-\yd)^3\mn^4\md - \yd(1-\yd)^4\mn^5 \bigg] \, .
\end{align}

The longitudinal distributions agree on the FS results
in the limit $Q^2 \rightarrow \infty$ except for a term
in Eq. (\ref{eqn:dld-1}). The factor $1/\yp{3} (1-\yd)^2$
is written as $1/\yp{2} (1-\yd)^3$ in Ref. \cite{fs}.
It is possibly a misprint \cite{fries}.



\end{document}